\PassOptionsToPackage{hyphens}{url}
\documentclass[sigconf,screen,nonacm]{acmart}
\setcopyright{none}

\usepackage{amsmath,amsfonts}
\usepackage{algorithmic}
\usepackage{graphicx}
\usepackage{textcomp}
\usepackage{url}
\usepackage{xcolor}
\usepackage{cleveref}
\usepackage{tikz}
\usetikzlibrary{calc, fit, positioning, arrows.meta, decorations.pathreplacing, backgrounds, matrix, patterns}
\usepackage{pgfplots}
\usepackage{pgfplotstable}
\pgfplotsset{compat=1.17}
\usepackage{subcaption}
\usepackage{enumitem}
\usepackage{booktabs}
\usepackage{threeparttable}
\usepackage{siunitx}
\usepackage{xspace}
\usepackage{bytefield}
\usepackage{listings}
\Crefname{lstlisting}{Listing}{Listings}
\usepackage{svg}
\usepackage{multirow}
\usepackage{makecell}
\definecolor{oiOrange}{RGB}{230,159,0}
\definecolor{oiSky}{RGB}{86,180,233}
\definecolor{oiGreen}{RGB}{0,158,115}
\definecolor{oiBlue}{RGB}{0,114,178}
\definecolor{oiVermillion}{RGB}{213,94,0}
\definecolor{oiPurple}{RGB}{204,121,167}
\definecolor{ink}{RGB}{30,30,30}
\AddToHook{env/figure/begin}{\let\footnotesize\small}
\AddToHook{env/figure*/begin}{\let\footnotesize\small}

\definecolor{rdaOp}{HTML}{1F4E79}      
\definecolor{rdaGroup}{HTML}{7F3F98}   
\definecolor{rdaField}{HTML}{005A32}   
\definecolor{rdaComment}{gray}{0.45}
\definecolor{rdaGroupField}{HTML}{B15928}  
\definecolor{rdaAnonOp}{HTML}{1F4E79}      
\definecolor{cmpnote}{gray}{0.4}

\lstdefinelanguage{rdaconfig}{
  morekeywords={
    Broadcast,AR_Dynamic,Bufferize,RepeatStatic, RepeatRef, AddressReader, GEMM, Matmul, Accum,
    BinaryMap,Accumulate,TiledReshape,AccumBigTile, mo1, mmo1, cu1, io1, mo2, mo3, mmo
  },
  morekeywords=[2]{Group,Meta,IOG,SMG,MMG,EWG},
  morekeywords=[3]{
    rank,data_in,data_out,ref_in,ref_out,repeat_cfg,parallel,size,is_left,is_top,count,mode,m,read_pattern,
    op,in_idx,out_idx,pulse_in,transposed,chunk,rows,cols, in, out, n, partial_in, partial_out, priority
  },
  morekeywords=[4]{
    targets, ready, placement, order
  },
  morekeywords=[5]{
    mo1, io1, cu1, mmo1
  },
  morekeywords=[6]{Group},
}

\lstdefinestyle{rdaconfig}{
  language=rdaconfig,
  basicstyle=\ttfamily\footnotesize,
  numberstyle=\footnotesize\color{gray},
  breaklines=true,
  columns=fullflexible,
  frame=single,
  captionpos=b,
  keywordstyle=\bfseries\color{rdaOp},
  keywordstyle=[2]\bfseries\color{rdaGroup},
  keywordstyle=[3]\color{rdaField},
  keywordstyle=[4]\color{rdaGroupField},
  keywordstyle=[5]\color{rdaAnonOp},
  comment=[l]{//},
  commentstyle=\itshape\color{rdaComment},
}

\def\BibTeX{{\rm B\kern-.05em{\sc i\kern-.025em b}\kern-.08em
    T\kern-.1667em\lower.7ex\hbox{E}\kern-.125emX}}

\definecolor{airedcolor}{HTML}{B3263E}

\definecolor{redundcolor}{HTML}{D97706}
\definecolor{redsugcolor}{HTML}{7C3AED}

\newcommand{\name}{\textsc{Elasticy}\xspace}
\newcommand{\nameCache}{\textsc{\name{}-Cache}\xspace}

\newcommand{\DLSM}{\textsc{DLSM}\xspace}

\newcommand{\nameLocal}{\textsc{\name{}-Local}\xspace}
\newcommand{\appType}{dynamic dense }

\newcommand{\VSOLong}{Virtual Stream Operator}
\newcommand{\VSO}{VSO}
\newcommand{\VSGLong}{Virtual Stream Group}
\newcommand{\VSG}{VSG}
\newcommand{\PSGLong}{Physical Stream Group}
\newcommand{\PSG}{PSG}
\newcommand{\PSOLong}{Physical Stream Operator}
\newcommand{\PSO}{PSO}
\newcommand{\Downtiling}{Retiling}

\begin{document}

\title{\name{}: A Runtime-Reconfigurable Architecture for Dynamic Streams}

\author{Konstantin Hossfeld}
\authornote{Equal contribution.}
\email{hossfeld@stanford.edu}
\affiliation{\institution{Stanford University}\city{Stanford}\state{California}\country{USA}}
\author{Nathan Sobotka}
\authornotemark[1]
\email{nsobotka@stanford.edu}
\affiliation{\institution{Stanford University}\city{Stanford}\state{California}\country{USA}}
\author{Gina Sohn}
\email{ginasohn@stanford.edu}
\affiliation{\institution{Stanford University}\city{Stanford}\state{California}\country{USA}}
\author{Tun-Yu Chang}
\email{tunyu@stanford.edu}
\affiliation{\institution{Stanford University}\city{Stanford}\state{California}\country{USA}}
\author{Kunle Olukotun}
\email{kunle@stanford.edu}
\affiliation{\institution{Stanford University}\city{Stanford}\state{California}\country{USA}}
\renewcommand{\shortauthors}{}

\begin{abstract}
    Reconfigurable Dataflow Accelerators (RDAs) have demonstrated high performance and energy efficiency for dense, regular tensor applications. However, conventional RDAs with static configurations show dramatically reduced performance and efficiency on irregular tensor workloads with data-dependent, non-uniform, dynamic tensor shapes, such as modern AI applications. To eliminate the deficiencies of conventional RDAs, we propose \name, a dynamically reconfigurable dataflow accelerator that employs streaming and tile abstractions to minimize the overhead of dynamic support. By encoding dynamic shapes into streams using rank-variable expressions, \name exploits data-dependent tensor shape dynamism to minimize off-chip data traffic and wasted compute cycles. \name employs stream-embedded reconfiguration information to trigger fine-grained runtime scheduling decisions. This capability improves performance by maximizing the dynamic spatiotemporal overlap between kernels. \name also introduces a novel distributed linked streaming memory (\DLSM) that more effectively uses on-chip memory to minimize off-chip data traffic and support dynamically sized tensors.

    Importantly, both reconfiguration and shape dynamism incur minimal overhead in \name: reconfiguration occurs infrequently at the granularity of a stream of tiles, while rank-variable expressions enable lightweight handling of dynamic dimensions. \name retains the high performance of conventional static RDAs while adding dynamic capabilities at a minimal cost. On a Llama 3-8B decode layer, \name achieves $\mathbf{92.6\%}$ of its roofline performance, compared to $\mathbf{69.7\%}$ on an H100. On evaluated GEMMs, the \DLSM reduces total off-chip traffic by up to $\mathbf{61.4\%}$ relative to the best evaluated cache configuration, with no additional burden of programmer-managed memory. \name's distributed runtime scheduler overlaps dependent kernels, speeding up kernel pairs by up to $\mathbf{1.42\times}$. Programs can also run in both space and time: each phase of a layer's graph is pipelined spatially, while phases fold onto the same fabric over time. Compared to a traditional spatial accelerator, \name achieves a $\mathbf{6.34\times}-\mathbf{31.67\times}$ improved area delay product on full layers.
    
\end{abstract}

\maketitle

\section{Introduction}

\begin{figure}[t]
    \centering
    \input{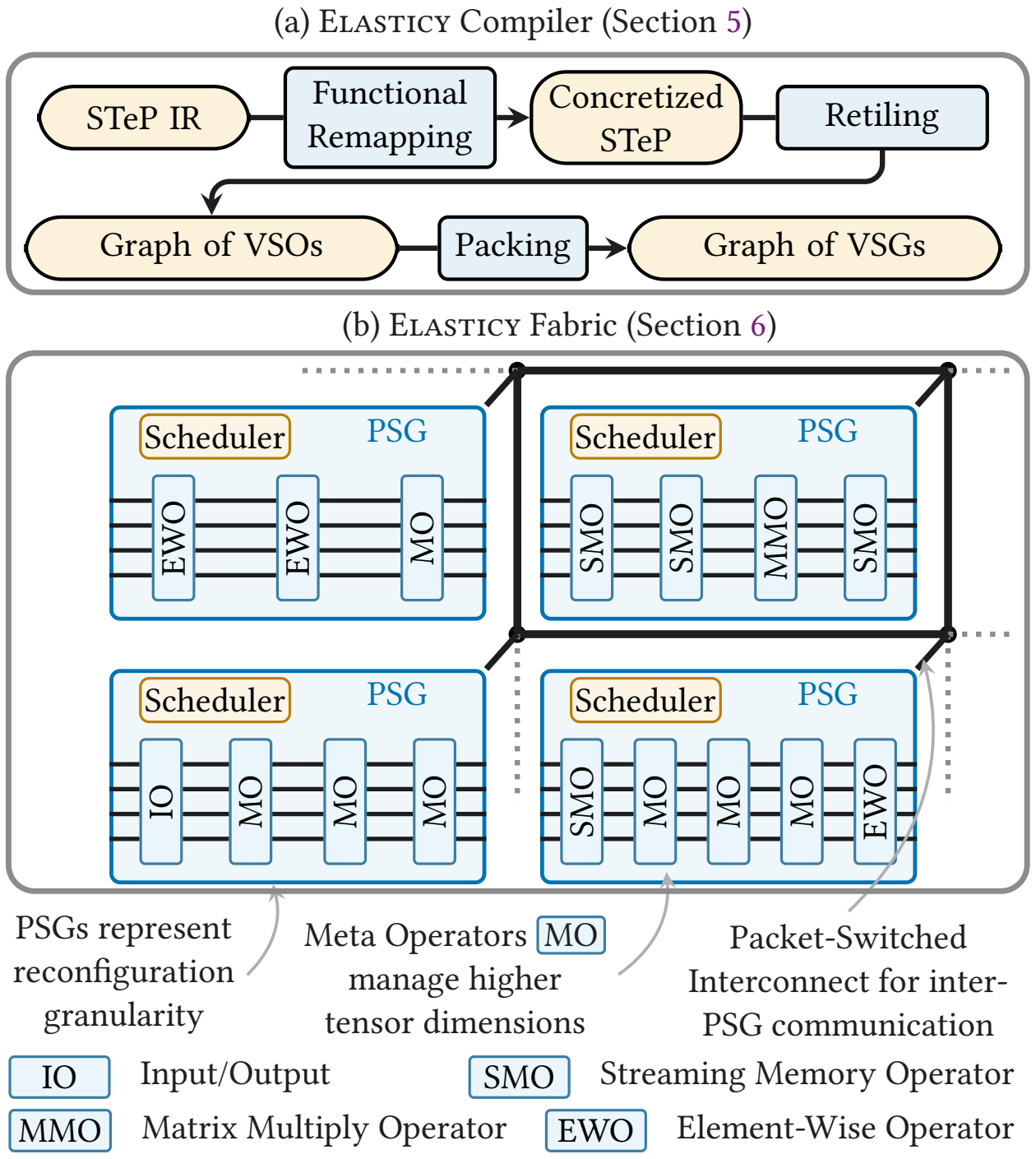}
    \caption{System Overview. (a) Compilation and Program Flow: the compiler maps applications described in the STeP IR to {\VSOLong}s ({\VSO}s), which are packed into {\VSGLong} ({\VSG}s) to form the bitstream~\cite{sohn2025step}. (b) The \name fabric and scheduler dynamically schedule the bitstream of {\VSG}s onto {\PSGLong}s ({\PSG}s).}
    \label{fig:fabric}
\end{figure}

Reconfigurable Dataflow Accelerators (RDAs) have emerged as a promising hardware accelerator paradigm. Their execution model enables seamless overlap of compute and memory operations, improving data reuse and reducing off-chip traffic through spatial pipelining.
Recent results suggest that RDAs can achieve competitive performance and high energy efficiency for several workloads, including LLMs~\cite{koeplinger2024kernelloopingeliminatingsynchronization, saadfalcon2025intelligencewattmeasuringintelligence, speculative-rda} and scientific computing~\cite{scientific-rda}.

However, the growing dynamism in many widely used tensor applications poses significant challenges to the memory systems and execution models of prior RDAs.

In modern ML workloads, for example, features such as data-dependent control flow~\cite{jiang2024mixtral,qwen3,openai2025gptoss120bgptoss20bmodel,glm4.5}, user-specific context~\cite{vllm}, and adaptive scheduling~\cite{yu2022orca, splitwise} introduce dynamically sized tensors that are difficult to efficiently support with existing scratchpad- or cache-based memory systems. 

Scratchpads must over-provision capacity for dynamically sized buffers since intermediate sizes are unknown at compile time, leading to fragmentation~\cite{ppa}. Cache-based architectures can handle these dynamically sized buffers; however, caches are reactive and do not handle stream access patterns and long-term look-ahead behavior efficiently. For streaming applications, caches introduce unnecessary memory traffic, conflict misses, and coherence protocol overheads~\cite{9407107}. Additionally, higher-level caches centralize off-chip traffic and can bottleneck throughput~\cite{10.1145/605397.605420}. High-throughput architectures, such as GPUs, expose scratchpads to avoid cache overheads for performance-critical code~\cite{dao2022flashattention, nvidia_cuda_memory}. An ideal memory system should handle dynamically sized buffers without introducing unnecessary off-chip traffic or putting an undue burden on software development.

Control flow and data-dependent tensor shapes change the balance of computation and data movement in a program~\cite{yang2025lserveefficientlongsequencellm}.
However, with conventional RDAs, resources are statically allocated and scheduled to each operator at compile time, limiting their ability to sustain efficient dataflow as runtime demands change.

This leaves two mapping choices. Mapping the whole graph spatially can leave memory and compute resources underutilized when the runtime demands differ from the compile-time assumption. Alternatively, dividing the program graph into bulk-synchronous phases exposes long pipeline fill and drain latencies, forces intermediate results off-chip, consuming unnecessary bandwidth, and prevents cross-phase prefetching.

To address these limitations, we present \name, a runtime-reconfigurable \appType dataflow accelerator that operates on streams of tiles. \name increases compute and memory utilization while decreasing off-chip traffic compared to a conventional RDA. 

\name's fabric, depicted in \Cref{fig:fabric}b, is composed of reconfigurable {\PSGLong}s ({\PSG}s), which contain {\PSOLong}s ({\PSO}s), such as IO, compute, and memory blocks, connected by a packet-switched credit-based on-chip network. Its distributed runtime scheduler dynamically loads {\VSGLong}s ({\VSG}s) onto {\PSG}s immediately after prior streams finish, sustaining high utilization without a static reconfiguration schedule. To support dynamically sized tensors, the memory operators implement a distributed linked streaming memory (\DLSM) that can spill data into neighbors, significantly reducing off-chip traffic over cache systems. \name additionally encodes tensor dimensions and reconfiguration triggers into data streams and operates on them through lightweight rank-variable operators (called meta operators), allowing for dynamic higher-order tensor shapes.

The \name compiler (\Cref{fig:fabric}a) maps applications written in the Streaming Tensor Programs (STeP)~\cite{sohn2025step} abstraction to our hardware. STeP is a tiled dataflow abstraction that enables \name to exploit abundant parallelism, support shape-independent programs, express control flow, and handle dynamically sized buffers.

\name makes the following contributions:
\begin{itemize}[leftmargin=*]
    \item A dynamic RDA with a stream-of-tiles execution model, enabling efficient execution of dense-matrix applications with dynamic, data-dependent irregularities.
    \item A lightweight distributed hardware scheduler that sustains high fabric utilization by embedding reconfiguration triggers in streams, eliminating global synchronization overhead and overlapping folded phases in time.
    \item A distributed linked streaming memory supporting streams of dynamically sized tensors and intermediate buffers, reducing off-chip traffic without programmer management.
    \item A set of lightweight hardware operators for tensor rank expressions to support high-performance execution of stream programs with tensor shape dynamism.
    \item A compiler that maps streaming dynamic tensor programs (STeP) onto \name, pipelining each phase of a graph spatially and folding the phases onto the same fabric across time.
\end{itemize}

We evaluate \name on a cycle-accurate Bluespec SystemVerilog simulator~\cite{Nikhil2008bluespec}, focusing on LLM decode and midfill workloads that exhibit dynamism through variable sequence lengths and data-dependent execution, including mixture-of-experts (MoE) routing. \name's evaluation demonstrates the benefits of runtime reconfiguration, distributed on-chip memory, and its lightweight support for tensor rank expressions. Across the evaluated LLM layers, \name achieves 75.2--96.5\% memory-roofline attainment.
Supporting ragged sequence lengths improves full-layer execution time by 1.15× relative to padded execution with the same placement and encoded instructions. On the evaluated GEMMs, distributed memory reduces total off-chip traffic by up to 61.4\% relative to the best evaluated cache configuration. Overlapping dependent kernels speeds up kernel pairs by up to $1.42\times$, and folding each layer's graph in time improves the area--delay product by up to $31.67\times$.


\section{Background}\label{sec:background}
We review prior work on reconfigurable dataflow accelerators and memory system designs for spatial architectures.

\subsection{Reconfigurable Dataflow Accelerators}


Dataflow applications are often mapped onto reconfigurable fabrics, including field-programmable gate arrays (FPGAs), coarse-grained reconfigurable architectures (CGRAs), and reconfigurable dataflow accelerators (RDAs)~\cite{prabhakar2018plasticine, umuroglu2017finn, zhang2021sara}. These architectures typically map dataflow graphs onto hardware resources at compile time and execute them as spatial pipelines.


However, static resource allocation can lead to pipeline underutilization when computational demands depend on runtime data. For example, in mixture-of-experts (MoE) LLM inference, token routing determines the number of tokens assigned to each expert, resulting in data-dependent batch sizes. These variations affect both computation and memory traffic, as well as their ratio (arithmetic intensity), due to differences in weight reuse~\cite{kim2026sieve}. 

The solution is to multiplex the fabric's resources based on runtime data. Previous work has categorized approaches based on whether they issue instructions dynamically in time (dynamic-issue) or space (dynamic-mapping)~\cite{chen2021elastic, anderson_cgra_me, ppa, parashar2013triggered, nguyen2021fifer}. Static-issue, static-mapping architectures generally cannot adapt to nonuniform, data-dependent compute demands. Therefore, they usually resort to the host system for making data-dependent decisions, introducing host communication and bulk-synchronous, full-system reconfiguration, and kernel launch overheads.

Dynamic-issue architectures aim to solve this. Instead of statically scheduling every operand within the bitstream, they encode the \textit{dependencies} between instructions and resolve them at runtime by executing only the instructions that are ready. Triggered Instructions~\cite{parashar2013triggered} dynamically select between a set of instructions within a compute core based on a set of predicates. FIFER~\cite{nguyen2021fifer} schedules at coarser-grained static data-rate pipelines instead of individual instructions, and selects instructions based on queue readiness. Both architectures employ static mapping, allowing a compile-time place \& route flow to minimize congestion. 


While there have been dataflow accelerators with more flexible architectures and execution models, they have not exploited the parallelism and compute density needed to accelerate dense-matrix applications with coarse-grained dynamism, which represents modern large-language models.

Dense applications with irregular components, such as dynamic tensor shapes, mix compute requirements with coarse-grained irregularities. We hypothesize that a reconfigurable static-mapping dynamic-issue fabric composed of large reconfigurable SIMD compute pipelines can yield dynamic scheduling benefits with minimal overhead, amortizing runtime decisions over long streams of tiles.

\subsection{Memory}
Static CGRAs and RDAs typically expose compiler- or programmer-managed scratchpad memories. Plasticine employs banked scratchpads in its Pattern Memory Units~\cite{prabhakar2018plasticine}, and SARA introduces compiler-managed memory consistency~\cite{zhang2021sara}. Scratchpads incur minimal hardware overhead. However, statically mapped scratchpads require storage placement and capacity to be provisioned ahead of execution. If the data size has a static upper bound, a program can be mapped but may incur memory fragmentation; if the data size is unbounded, off-chip memory traffic is required once buffers exceed the static partitioning. To address this, prior work has proposed configurable memory systems that create larger logical memories by sharing scratchpads. However, this compile-time decision incurs additional load latency for all accesses and still requires prior knowledge of data sizes~\cite{ppa}. On GPUs, PagedAttention addresses off-chip fragmentation, but on-chip fragmentation remains an issue~\cite{vllm}.

Alternatively, distributed architectures such as FIFER and TIA use private L1 caches and a shared last-level cache, as GPUs and multicore CPUs do~\cite{nguyen2021fifer, parashar2013triggered}. Caches naturally accommodate dynamically sized data without static partitioning. However, replacement policies lack stream lifetimes and future reuse information, so streaming applications thrash and cannot use large look-ahead windows to overlap compute with memory accesses~\cite{9407107}. Stream Floating addresses this by exposing stream access patterns through the ISA to enable proactive data movement, but introduces additional coherence and consistency overheads, as well as a tradeoff between prefetching aggressiveness versus accuracy and timeliness~\cite{9407107}. Additionally, centralized higher level memories contend through shared busses, hash collisions, and bank conflicts, a poor match for spatial, distributed architectures. To overcome these issues, GPUs shift on-chip memory management to software: FlashAttention explicitly tiles attention into SRAM~\cite{dao2022flashattention}, CUDA provides access-policy windows that preferentially retain selected data in L2~\cite{nvidia_cuda_l2}, and Hopper exposes programmer-managed distributed shared memory within a thread block cluster~\cite{nvidia_cuda_dsm}.

Non-Uniform Cache Architectures (NUCA) distribute cache banks across the chip, with access latency depending on data proximity. Static NUCA systems fix the placement of cache lines, while Dynamic NUCA and Reactive NUCA reduce latency by migrating or replicating data \cite{10.1145/605397.605420, hardavellas2009rnuca, merino2008spnuca}. We take inspiration from the non-uniform access latencies and block migration for our memory system. However, static NUCA systems do not exhibit ideal locality, while dynamic NUCA cache systems still have coherence and consistency overheads. 

Dynamic dense workloads require a memory system that combines flexible allocation with explicit streaming behavior. A memory architecture that supports dynamically sized buffers with low fragmentation while preserving high compute–memory overlap would significantly improve performance on dynamic dense workloads.

\section{System Overview}
\Cref{fig:fabric} provides an overview of \name, which consists of a compiler and a runtime-reconfigurable fabric. Programs written in The STeP programming abstraction (\Cref{sec:step}) are compiled into {\VSGLong}s ({\VSG}s) (\Cref{sec:compiler}), which serve as \name's ISA. At runtime, the fabric schedules {\VSG}s onto {\PSGLong}s ({\PSG}s) (\Cref{sec:fabric}). \name's fabric supports unbounded and dynamically sized intermediate buffers through its distributed linked streaming memory, and includes meta operators for handling dynamic tensor dimensions and control flow. 

\section{Frontend: Streaming Tensor Programs}\label{sec:step}

We adopt Streaming Tensor Programs (STeP) as \name's frontend IR~\cite{sohn2025step}. STeP represents tensor computations as graphs operating on streams of tiles, with explicit buffering and data-dependent control flow. Its shape-agnostic operators allow the same graph to process different runtime tensor shapes, while its tile representation matches the fabric's compute granularity. Explicit buffering exposes opportunities for reuse without requiring intermediate tensor sizes to be known at compile time. 


STeP encodes tensor shapes within streams using rank variables (i.e., \textbf{stop tokens}).
Each tensor dimension terminates with a stop token $S_N$ ($N \ge 1$), where $N$ indicates the rank of that dimension (e.g. $S_1$ marks the end of a vector).
For example, two \textit{ragged} scalar tensors are expressed as follows:
\[
1\ 2\ 3\ \textbf{S1}\ 4\ 5\ \textbf{S2}\
11\ 12\ 13\ \textbf{S1}\ 14\ 15\ 16\ \textbf{S2}\ \textbf{end}
\]
These boundaries allow operators to process dynamically sized and ragged tensors without knowing their dimensions in advance. STeP provides rank-variable operators that execute on these streams, including off-chip memory, on-chip memory, higher-order, control-flow, and shape operators~\cite{sohn2025step}. \Cref{tab:step-operators} summarizes a subset of the operators used in \name's STeP-based representation. We refer readers to the STeP paper~\cite{sohn2025step} for the full operator list and semantics.

\begin{table}[t]
\caption{Selected structural and control-flow operators in \name's STeP-based representation~\cite{sohn2025step}.}
\label{tab:step-operators}
\centering
\begingroup
\footnotesize
\setlength{\tabcolsep}{3pt}
\begin{tabular}{ll}
\toprule
\textbf{Operator} & \textbf{Effect} \\
\midrule
AccumRetile & Accumulate a stream into a larger output tile. \\
Flatten & Eliminate a rank from input stream \\
Promote & Increase rank / introduce tiling \\
RepeatStatic & Repeat input data a fixed number of times \\
RepeatRef & Repeat input data a dynamic number of times \\
LogicalTiledReshape & Reshape tile grid (no data change) \\
LogicalRetileStreamify & Retile then emit as stream with stop tokens \\
\midrule
Broadcast & Duplicate a stream to N branches \\
Partition & route data to dynamically selected outputs \\
Reassemble & merge selected inputs into one stream \\
\bottomrule
\end{tabular}
\endgroup
\end{table}

When compiling STeP to \name, control is embedded into streams by annotating each tile with a stop token and, optionally, a reconfiguration trigger (i.e. an \textbf{end token}). Since tiles are large, the overhead of encoding dynamic shapes into the data stream amortizes well ($<1\%$).
We translate STeP to our accelerator with a compiler that maps STeP programs with dynamic shapes and data-dependent control flow to our fixed-size hardware. Unlike loop-centric compiler IRs, STeP makes streaming structure, dynamic shapes, and control flow information explicit, making it a natural frontend for \name~\cite{llvm-mlir-linalg, chen2018tvm, sohn2025step}.

\Cref{tab:step-elasticy-comparison} summarizes the division of responsibilities between STeP and \name's extensions to the STeP framework.

\begin{table}[h]
\centering
\caption{Component contributions: STeP vs.\ \name}
\begingroup
\footnotesize
\setlength{\tabcolsep}{3pt}
\begin{tabular}{lcc@{}}
\toprule
\textbf{Component} & \textbf{STeP} & \textbf{\name} \\
\midrule
Programming Abstraction & Yes & Uses it \\
Bufferize Semantics & Yes & Uses them \\
IR-level meta-op semantics & Yes & Uses them \\
Hardware Architecture & No & Yes \\
Stream-Embedded Reconfiguration & No & Yes \\
DLSM Memory System & No & Yes \\
Rank-Variable Operators & Yes & Uses them\\ 
(Meta-Op) Hardware & No & Yes \\
Runtime Scheduler & No & Yes \\
Backend Compiler & No & Yes \\
\bottomrule
\end{tabular}
\endgroup
\label{tab:step-elasticy-comparison}
\end{table}

\begin{figure*}[!t]
\centering
\makebox[\textwidth][c]{%
  \begin{minipage}[c]{.68\textwidth}
    \par\smallskip
\definecolor{stepExpandTint}{RGB}{172,96,148}

\begin{tikzpicture}[
x=0.825cm, y=0.70cm,
    >=stealth,
    every node/.style={font=\footnotesize, align=center},
    box/.style={draw=ink!45, line width=0.75pt, rounded corners=2pt,
                minimum width=1.5cm, minimum height=0.5cm},
    orangeNode/.style={box, draw=oiOrange, line width=0.8pt, fill=oiOrange!22, text=ink},
    goldNode/.style={box, draw=oiBlue, line width=0.8pt, fill=oiBlue!24, text=ink},
    repeatRefNode/.style={box, draw=oiSky, line width=0.8pt, fill=oiSky!24, text=ink},
    repeatRefElided/.style={box, draw=ink, line width=0.75pt, densely dotted, fill=ink!20, text=ink},
    plumNode/.style={box, draw=oiPurple, line width=0.8pt, fill=oiPurple!20, text=ink},
    seafoamNode/.style={box, draw=oiGreen, line width=0.8pt, fill=oiGreen!22, text=ink,
      postaction={pattern=north east lines, pattern color=ink!65, opacity=.35, draw=none}},
    pinkNode/.style={box, draw=oiVermillion, line width=0.8pt, fill=oiVermillion!20, text=ink},
    blueNode/.style={box, draw=oiBlue, line width=0.8pt, fill=oiBlue!24, text=ink},
    redNode/.style={box, draw=oiVermillion, line width=0.8pt, fill=oiVermillion!25, text=ink},
    reassembleNode/.style={box, draw=oiBlue, line width=0.8pt, fill=oiBlue!28, text=ink},
    controlNode/.style={box, draw=oiOrange, line width=0.8pt, fill=oiOrange!25, text=ink},
    whiteNode/.style={box, draw=ink!50, line width=0.75pt, fill=white, text=ink},
    elidedSubgraph/.style={box, draw=black, densely dotted, line width=0.75pt,
                           fill=white, text=ink},
    flow/.style={->, line width=0.75pt, draw=ink!55},
    remember picture
]

\tikzset{flow/.style={-{Stealth[length=2.1mm,width=1.3mm]},line width=.75pt,draw=ink!70},
 refFlow/.style={flow,densely dashed},refLine/.style={draw=ink!70,line width=.75pt,densely dashed}}
\node[whiteNode] (larA) at (-4.2,2.3) {Address Reader};
\node[whiteNode] (lbc) at (-.5,2.3) {Broadcast};
\node[whiteNode] (larB) at (-4.2,1.1) {Address Reader};
\node[whiteNode] (larC) at (-.5,1.1) {Address Reader};
\node[whiteNode] (lab) at (-4.2,-.3) {Map: Matmul$^{\textstyle\mathrm{T}}$};
\node[whiteNode] (lac) at (-.5,-.3) {Map: Matmul$^{\textstyle\mathrm{T}}$};
\node[whiteNode] (lqk) at (-4.2,-1.95) {Map: Matmul$^{\textstyle\mathrm{T}}$};
\node[whiteNode] (lfinal) at (-.5,-3.16) {Map: Matmul$^{\textstyle\mathrm{T}}$};
\node[whiteNode] (larD) at (-.5,-1.82) {Address Reader};
\draw[flow] (larA.east) -- node[midway,above,yshift=-2pt] {$A_{\textstyle i}$} (lbc.west);
\draw[flow] (lbc.south) -- (-.5,1.75) -- (-2.5,1.75) -- (-2.5,-.3) -- (lab.east);
\draw[flow] (lbc.east) -- (1.2,2.3) -- (1.2,-.3) -- (lac.east);
\draw[flow] (larB.south) -- node[midway,left] {$B_{\textstyle i}^{\textstyle\top}$} (lab.north);
\draw[flow] (larC.south) -- node[midway,left] {$C_{\textstyle i}^{\textstyle\top}$} (lac.north);
\draw[flow] (lab.south) -- node[midway,left] {$A_{\textstyle i}B_{\textstyle i}$} (lqk.north);
\draw[flow] (lac.south) -- (-.5,-1.25) -- node[pos=.68,above] {$A_{\textstyle i}C_{\textstyle i}$} (-3.78,-1.25) -- ([xshift=.3465cm]lqk.north);
\draw[flow] (lqk.south) |- node[pos=.75,above] {$A_{\textstyle i}B_{\textstyle i}(A_{\textstyle i}C_{\textstyle i})^{\textstyle\top}$} (lfinal.west);
\draw[flow] (larD.south) -- node[midway,right] {$D_{\textstyle i}^{\textstyle\top}$} (lfinal.north);
\draw[flow] (lfinal.south) -- node[midway,left,yshift=-3pt] {$Y_{\textstyle i}$} (-.5,-3.85);
\draw[draw=ink!60,line width=.8pt,rounded corners=5pt]
 (-5.65,-4.14) rectangle (1.35,2.75);
\node[font=\footnotesize\bfseries] at (-2.15,3.05) {STeP Computation};
\begin{scope}[on background layer]
\node[fit=(larC)(lac),inner sep=5pt,rounded corners=5pt,
 draw=stepExpandTint,line width=.85pt,densely dotted,fill=stepExpandTint!12] (acSeed) {};
\end{scope}
\draw[stepExpandTint,line width=.8pt,densely dotted,-{Stealth[length=2.1mm,width=1.3mm]}]
 (acSeed.north east) -- (1.85,2.15);
\draw[stepExpandTint,line width=.8pt,densely dotted,-{Stealth[length=2.1mm,width=1.3mm]}]
 (acSeed.south east) -- (1.85,-3.44);
\begin{scope}[shift={(3.38cm,0cm)}]
\node[pinkNode] (arC) at (3,2.3) {Address Reader};
\node[seafoamNode] (bufA) at (0,.5) {\texttt{\bfseries smo0}\\Bufferize\\Rank=1};
\node[seafoamNode] (bufC) at (3,.5) {\texttt{\bfseries smo1}\\Bufferize\\Rank=2};
\node[plumNode] (bcA) at (0,-.85) {Broadcast};
\node[plumNode] (rrA) at (0,-2.1) {RepeatStatic\\Count=$N$};
\node[orangeNode] (rrC) at (3,-1.92) {RepeatRef};
\node[seafoamNode] (mm) at (0,-3.44) {\texttt{\bfseries mmo0}\\Matmul$^{\textstyle\top}$};
\node[seafoamNode] (acc) at (2.5,-3.44) {\texttt{\bfseries ewo0}\\Accum};
\draw[flow] (0,2.75) -- (0,1.8) -| ([xshift=.3cm]bufA.north);
\draw[flow] (arC.south) -- (3,1.8) -| ([xshift=-.3cm]bufC.north);
\node[anchor=east] at (-.08,2.2) {$A_{\textstyle i}$};
\node[anchor=west] at (2,1.6) {$C_{\textstyle i}^{\textstyle\top}$};
\draw[refFlow] (bufA.south) -- (bcA.north);
\draw[refFlow] (bcA.south) -- (rrA.north);
\draw[refFlow] (bcA.east) -- (3,-.85) -- (rrC.north);
\draw[refFlow] (bufC.east) -- (4.2,.5) -- (4.2,-1.92) -- (rrC.east);
\draw[refFlow] (rrA.west) -- (-1.2,-2.1) -- (-1.2,1.7) -| ([xshift=-.3cm]bufA.north);
\draw[refFlow] (rrC.south) -- (3,-2.62) -- (4.65,-2.62) -- (4.65,1.7) -| ([xshift=.3cm]bufC.north);
\draw[flow] (bufA.west) -- (-1.75,0.5) -- (-1.75,-3.44) -- (mm.west);
\draw[flow] (bufC.west) -- (1.35,0.5) -- (1.3,-2.75) -| (mm.north);
\draw[flow] (mm.east) -- (acc.west);
\draw[flow] (acc.east) -- node[above] {$A_{\textstyle i}C_{\textstyle i}$} (4.6,-3.44);
\end{scope}
\draw[draw=ink!60,line width=.8pt,rounded corners=6pt] (1.85,-4.14) rectangle (8.9,2.75);
\node[font=\footnotesize\bfseries] at (5.375,3.05) {VSG Graph ($A_{\textstyle i}C_{\textstyle i}$)};
\end{tikzpicture}

    \par\smallskip
    \input{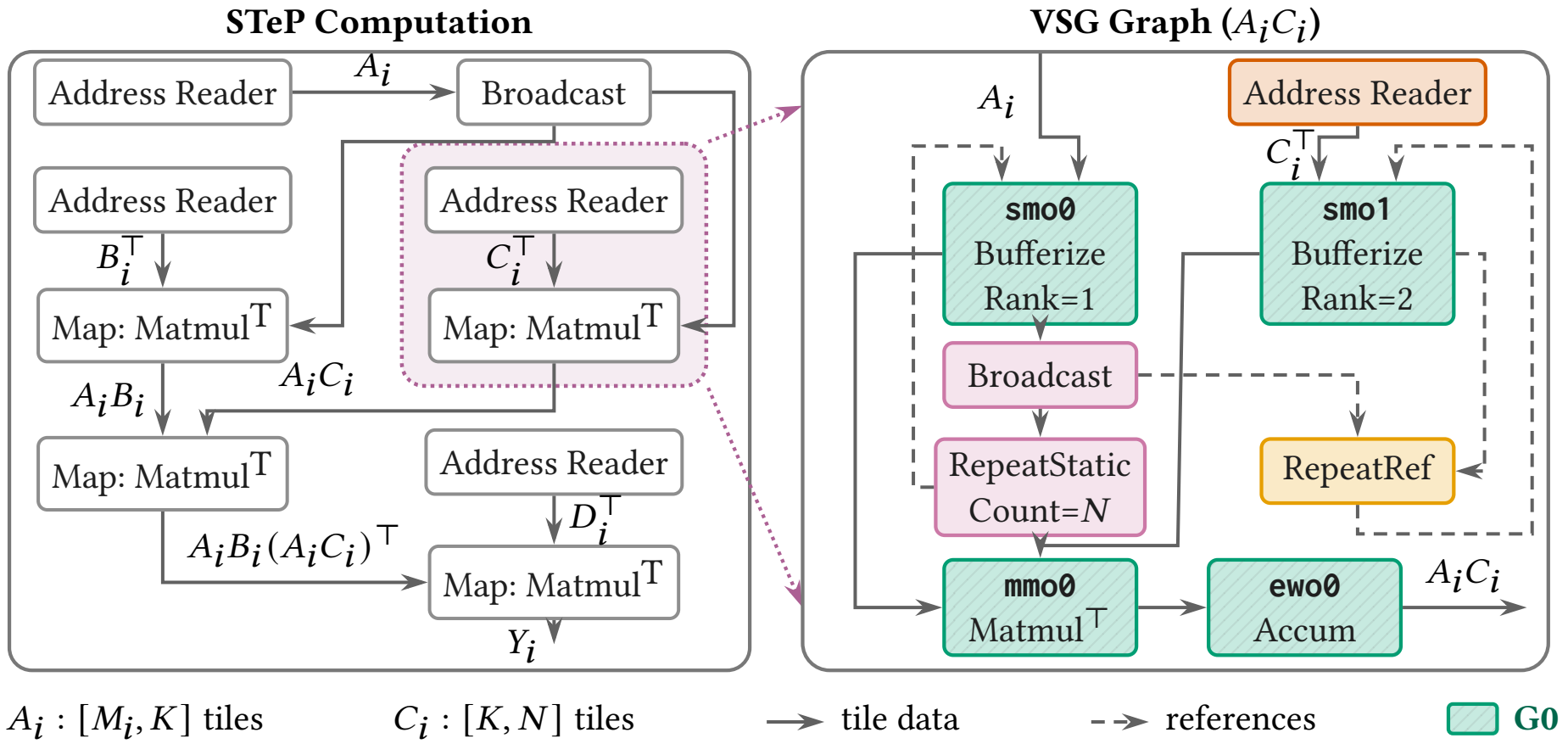}
    \caption{STeP graph for $A_{\text{\footnotesize i}}B_{\text{\footnotesize i}}(A_{\text{\footnotesize i}}C_{\text{\footnotesize i}})^{\text{\footnotesize$\top$}} D_{\text{\footnotesize i}}$, with the highlighted matmul expanded into a {\VSG} graph. The row dimension of $A_{\text{\footnotesize i}}$ is dynamic and the column dimension of $C_{\text{\footnotesize i}}$ is static. Colors distinguish {\VSG}s; hatched operators form G0, shown in \Cref{lst:isa-example}.}
        \label{fig:STeP Graph}
  \end{minipage}%
  \hfill
  \begin{minipage}[c]{.29\textwidth}
    \par\smallskip
\begin{lstlisting}[style=rdaconfig,literate={G0}{{\bfseries\color{oiGreen!65!black}G0}}2 {G1}{{\bfseries\color{oiOrange!65!black}G1}}2,caption={Selected ISA fields for G0 in \Cref{fig:STeP Graph}.},label={lst:isa-example}]
Group G0 : MMG
  smo0 = Bufferize { rank=1,
    data_in=0, data_out=0,
    ref_in=1, ref_out=1 }
  smo1 = Bufferize { rank=2,
    data_in=3, data_out=2,
    ref_in=4, ref_out=3 }
  mmo0 = Matmul {data_in=[0,2], 
    data_out=0}
  ewo0 = Accum {data_in = 0, 
    data_out = 0, rank = 1}
  targets = [...]
  order = 0
  placement = (1, 1)
\end{lstlisting}
    \par\smallskip
  \end{minipage}%
}%
\end{figure*}

\section{Compilation}\label{sec:compiler}
This section explains how to translate STeP, which operates on tiles of arbitrary size, to a graph of {\VSGLong} ({\VSG}s), which are placed onto {\PSGLong}s ({{\PSG}s)~\cite{sohn2025step}. An input STeP program can be annotated with candidate points for \textit{graph folding}, which implies which parts of the graph the compiler \textit{may} execute serially in time. The resulting ISA contains placed and routed {\VSG}s, and leaves temporal stream-level scheduling and storage-occupancy decisions to runtime. 

\subsection{Functional Remapping}
The functional remapping stage canonicalizes the STeP IR: n-ary partition, broadcast, and reassemble are rewritten into balanced trees with fixed fan-in and fan-out. Map and Accumulate functions are decomposed down to the fabric's physical operators.

\subsection{\Downtiling}\label{subsec:downtiling}
After canonicalization, the STeP program is expressed over compile-time fixed logical tiles, even though its higher-order stream shape remains fully dynamic. Logical tiles determine the granularity of off-chip data movement but need not match the physical tile size of the fabric. The {\Downtiling} pass hierarchically tiles each logical tile by introducing two innermost stream dimensions indexing its physical-tile grid, leaving the outer dynamic stream structure unchanged.

For element-wise, shape operators, and intermediate buffers, each operator is lifted over the intra-tile dimensions, incrementing rank-sensitive arguments by two. By construction, the relative ranks of the corresponding reference streams are preserved, and expressions that depend on reference streams are unchanged. 

Operators with non-element-wise semantics, such as matrix multiplication, require a blocked lowering over the physical-tile grid. For a matmul, the pass uses Bufferize and Repeat operators to replay the operands along the appropriate tile dimensions and make intra-tile reuse explicit, as demonstrated in~\Cref{fig:STeP Graph}.

\subsection{Packing}
We pack the resulting program's {\VSO}s into groups of PSGs, obeying group type requirements. Tighter groups allow more operations to fit onto the fabric simultaneously and reduce reconfigurations~\cite{prabhakar2018plasticine, nguyen2021fifer}. For matrix multiplications, packing co-locates operand replay, multiplication, and accumulation into a single {\PSG}, removing the lowest-level, retiling-introduced operand replay traffic from the inter-{\PSG} interconnect. {\VSG}s are represented by colors in~\Cref{fig:STeP Graph}.

\subsection{Separation into Contexts \& Placement}
Next, the fully spatial graph must be folded onto the fabric, by partitioning the {\VSG}s into a tree of subgraphs called \emph{contexts}. Context boundaries are delineated using higher-rank bufferization operations. Each context must fit within the on-chip resources; once the contained {\VSG}s are scheduled, they execute to completion.

For each context $\kappa$, we statically place its {\VSG}s using simulated annealing. Formally, a placement is a mapping $\pi : \mathrm{VSG} \to {PSG}$ such that for ${\VSG}_1, {\VSG}_2 \in \kappa$, $\pi({\VSG}_1) \neq \pi({\VSG}_2)$. Intermediate buffer {\VSG}s that connect two contexts are members of both contexts and must be located on the same {\PSG}. Our placement objective penalizes shared NoC routes within contexts using wire length as a tie-breaker. If it is not possible to separate the program into sufficiently small contexts, we return an error.

\subsection{ISA}\label{sec:isa}
\Cref{lst:isa-example} shows an excerpt of a program expressed in the \name ISA. The program is encoded as a set of \textcolor{rdaGroup}{\texttt{\textbf{Groups}}} ({\VSG}s), each containing an ordered series of \textcolor{rdaOp}{\texttt{\textbf{Operator}}} ({\VSO}) declarations with explicit \textcolor{rdaField}{\texttt{\textbf{input\_port}}} and \textcolor{rdaField}{\texttt{\textbf{output\_port}}} mappings and {\VSO}-specific parameters (e.g. \textcolor{rdaField}{\texttt{\textbf{rank}}}). Each \textcolor{rdaGroup}{\texttt{\textbf{Group}}} also specifies \textcolor{rdaGroupField}{\texttt{\textbf{Placement}}}, which assigns the {\VSG} to a {\PSG}, and output \textcolor{rdaGroupField}{\texttt{\textbf{Targets}}}, which identify each destination's physical location, local {\VSG} configuration slot, and input port. The \textcolor{rdaGroupField}{\texttt{\textbf{Ready}}} flag indicates a {\VSG} is scheduled immediately upon configuration, otherwise, each \textcolor{rdaGroup}{\texttt{\textbf{Group}}} is enabled by input availability. An additional \textcolor{rdaGroupField}{\texttt{\textbf{Order}}} field can additionally be used to constrain activation order among {\VSG}s assigned to the same {\PSG}.


\section{Reconfigurable Fabric}\label{sec:fabric}

The \name fabric (\Cref{fig:fabric}b) consists of {\PSOLong}s ({\PSO}s), which fire on input availability. They are grouped into heterogeneous {\PSGLong}s ({\PSG}s). The {\PSG}s are connected by a credit-based, packet-switched, mesh interconnect. The ISA's {\VSGLong}s ({\VSG}s) are dynamically scheduled onto {\PSG}s using distributed schedulers.

\subsection{{\PSGLong}s}
We adopt the two-tiered interconnect paradigm from Plasticine~\cite{prabhakar2018plasticine}. Individual operators are sequentially ordered in space within {\PSG}s, and {\PSG}s communicate through the interconnect. When a {\PSO} in a {\PSG} does not select an input port, its data are forwarded to the next {\PSO}. This reduces interconnect complexity and keeps lowest-tiling-level traffic local to a single {\PSG} for GEMMs. By reconfiguring at the {\PSG} granularity, this also reduces scheduler invocations.

\subsection{Scheduler}

\begin{figure}[ht!]
\begin{tikzpicture}[x=.99cm,font=\footnotesize\rmfamily,
 shell/.style={draw=black!60,line width=.5pt,fill=oiBlue!6,rounded corners=2pt},
 block/.style={draw=black!55,line width=.45pt,fill=white,rounded corners=1.5pt,
   align=center,inner xsep=3pt,inner ysep=1pt,outer sep=0pt,minimum width=2.05cm,minimum height=.34cm},
 scheduler/.style={block,fill=oiGreen!14},
 flow/.style={-stealth,draw=black!75,line width=.55pt},
 lab/.style={font=\footnotesize\rmfamily,align=center,text=black!85,inner sep=1.5pt}]
\draw[shell] (0,.08) rectangle (2.35,1.50);
\draw[shell] (4.25,.08) rectangle (8.5,1.50);
\node[font=\footnotesize\rmfamily\bfseries] at (1.175,1.31) {Source PSG};
\node[font=\footnotesize\rmfamily\bfseries] at (6.85,1.31) {Target PSG};
\node[block] (source) at (1.175,1.00) {VSG $s$};
\node[scheduler] at (1.175,.30) {Local scheduler};
\node[scheduler] (sched) at (6.85,1.00) {Local scheduler};
\node[block,fill=oiOrange!10,minimum width=.85cm] (buffer) at (4.85,1.00) {$t$};
\node[lab,anchor=center] at (4.85,1.31) {Buffer};
\draw[black!50,line width=.4pt] ([xshift=-.12cm]buffer.north) -- ([xshift=-.12cm]buffer.south);
\draw[black!50,line width=.4pt] ([xshift=-.25cm]buffer.north) -- ([xshift=-.25cm]buffer.south);
\draw[flow] (buffer.east) -- (sched.west);
\node[block,minimum width=1.85cm,dashed,draw=black!45,fill=black!4,text=black!65] (old) at (7.35,.30) {Previous VSG};
\node[block,minimum width=1.85cm] (target) at (5.35,.30) {VSG $t$};
\draw[flow] (old.north) -- ([xshift=.495cm]sched.south);
\node[lab,anchor=west] at (7.40,.65) {Done};
\draw[flow] ([xshift=-.099cm]sched.south) |- (5.35,.53) -- (target.north);
\node[lab,anchor=south,inner ysep=.35pt] at (6.00,.53) {\textbf{2} Configure};
\draw[flow] (source.east) -- (buffer.west);
\node[lab,anchor=south] at (3.30,1.02) {\textbf{1} Request $t$};
\draw[flow] ([yshift=.109cm]target.west) -- (2.35,.41);
\node[lab,anchor=south] at (3.30,.43) {\textbf{3} Ready};
\draw[flow,line width=.95pt] (2.35,.12) -- (4.25,.12);
\node[lab,anchor=south] at (3.30,.14) {\textbf{4} Stream data};
\end{tikzpicture}
    \caption{Connection establishment with distributed scheduling. (1) The source requests a connection to {\VSG} $t$ at its compiler-assigned {\PSG}; the destination buffers the request. (2) After the previous configuration completes and drains, the local scheduler configures $t$, subject to compiler-specified \textcolor{rdaGroupField}{\texttt{\textbf{Order}}} constraints. (3) The destination acknowledges readiness. (4) The source begins transmitting stream data.}
    \label{fig:scheduling}
\end{figure}
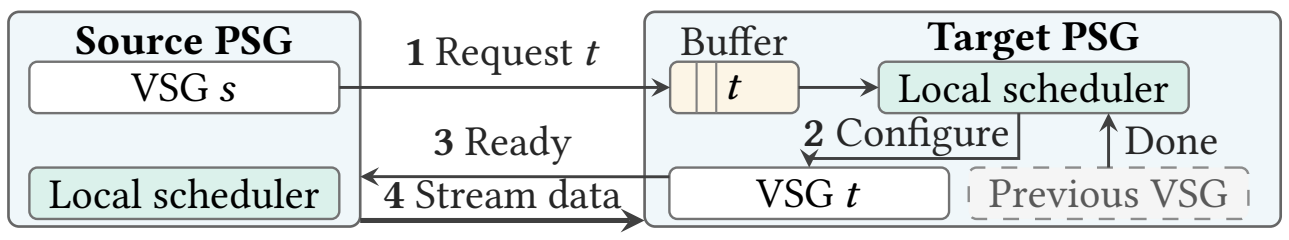

The scheduler sustains high fabric utilization by configuring idle {\PSG}s to run ready {\VSG}s and establishing connections between scheduled {\VSG}s, as depicted in~\Cref{fig:scheduling}.

\noindent\textbf{Allocating {\VSG}s and Establishing Connections.}
Each {\PSG} contains a local scheduler that stores the configurations of its compiler-assigned {\VSG}s. An active source {\VSG} $s$ requests a connection to a target {\VSG} $t$ at its compiler-specified physical location. The destination scheduler buffers the request until its {\PSG} is available and the compiler-specified execution \textcolor{rdaGroupField}{\texttt{\textbf{Order}}} permits $t$ to run. Once the {\VSG} is configured, $t$ acknowledges the request, allowing $s$ to transmit data. This protocol prevents data from entering the interconnect before the destination is ready to receive it.


\noindent\textbf{Deallocation.} 
A {\VSG} finishes after its input streams have ended and its output data and end tokens have drained. The {\PSG} signals completion to its local scheduler, releasing the resource for another eligible {\VSG} assigned to that location.

Importantly, because scheduling occurs only at stream boundaries, scheduler overhead is amortized over long steady-state executions. Across the evaluated LLM layers, the fabric applies one configuration per approximately 500-5,200 execution cycles.

\subsection{Distributed Linked Streaming Memory (\DLSM)}

\begin{figure*}[t]
    \centering
    \input{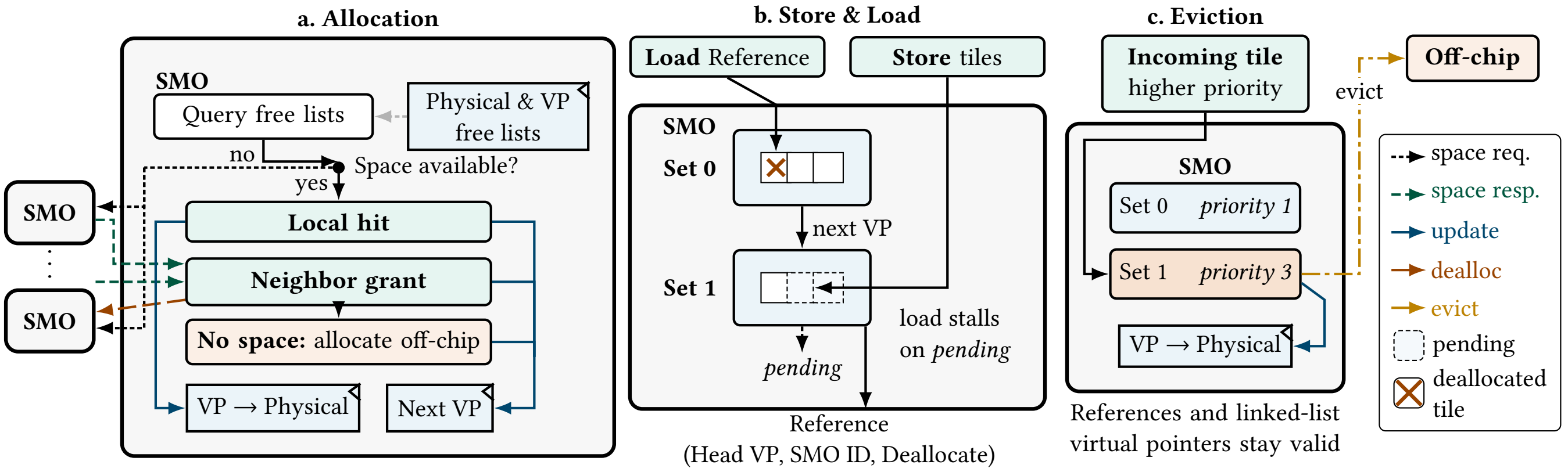}
    \caption{\name's \DLSM. (a) Allocation proceeds local-first, then via one-hop neighbor requests, and finally off-chip. (b) Stores append tiles into fixed-size linked sets and emit a virtual reference to the stream head; loads later traverse the linked list. (c) Eviction preserves reference validity by moving data off-chip and updating only the virtual-to-physical mapping.}
    \label{fig:memory-system}
\end{figure*}

STeP~\cite{sohn2025step} provides a Bufferize operation to store dynamically-sized write-once, read-many streams, delineated by stop tokens greater than or equal to the bufferize rank. Bufferize operators emit references, and reads are triggered by re-inputting the reference.

\name implements STeP's semantics with a novel memory system: a distributed linked streaming memory (\DLSM), which is implemented over a mesh of Streaming Memory Operators (SMOs). The {\DLSM} objectives are threefold: support dynamically sized buffers, achieve high utilization of on-chip capacity, and enable high throughput for all operating modes.

\noindent\textbf{Memory Organization}
Each stream is stored as a spatially distributed linked list of sets, each containing a fixed number of tile frames. Each frame carries next-link metadata indicating valid, pending, or stream termination. Each set additionally contains a virtual pointer (VP) to the next set and its owning SMO, resolved through an SMO-specific virtual-to-physical table. For the evaluated fabric, metadata occupies $0.11\%$ of the stored data capacity.
The load and store paths are independent and operate in parallel.

\noindent\textbf{Distributed Storage Allocation}
The storage allocation process is shown in \Cref{fig:memory-system}a. To hide latency, storage is eagerly allocated. Allocation proceeds local-first, then neighbors, and finally off-chip. If neighbors allocate space, the requester accepts the first response and deallocates unused offers. SMOs periodically repoll for more desirable (physically closer) storage locations.

Message passing occurs over a lightweight mesh signal interconnect, while data is carried on the tile-width NoC. Allocating at a set granularity greatly reduces memory-management communication and enables load and store pipelining.

\noindent\textbf{Stores and References}
\Cref{fig:memory-system}b depicts the storage process. Incoming tiles are forwarded to their pre-allocated location. Appending a tile updates the preceding tile's next-link metadata.

Once a stream is stored, the SMO emits a reference, consisting of the head VP of the stream, the ID of the SMO that produced the reference, and a deallocation flag set to true. Operators such as repeat nodes that duplicate references rewrite the deallocate bit so that only the final consumer is destructive. This promptly reclaims storage without software management or garbage collection.

\noindent\textbf{Loading Data}
Loads are depicted in \Cref{fig:memory-system}b with virtual-to-physical mappings omitted. A load is issued by enqueuing a reference into an SMO and traversing the associated linked list. Loads stall at frames whose next-link is marked as \textit{pending}. Non-destructive loads leave the linked list intact. Upon completion of a destructive read of a set, the set and associated VP are deallocated.

\noindent\textbf{Priority Eviction}
When on-chip space is exhausted, an SMO can evict a lower-priority set instead of spilling incoming data (\Cref{fig:memory-system}c). Eviction updates
the set's virtual-to-physical mapping, preserving existing
references and links. Priorities are configurable; by default, higher-rank buffers are evicted first.



\noindent\textbf{Memory Modes}
Beyond sequential replay, SMOs support strided traversal,
shuffle, accumulation, and passthrough modes. Strided traversal uses multiple saved head pointers to read stored tiles in prescribed strided orders. Shuffle appends incoming items to buckets selected by an accompanying index stream and then emits the buckets in order. Accumulation sums stored partials with input tiles, using an in-SMO, add-only accumulation unit. Passthrough automatically drains arriving tiles while preserving stream order, providing elastic buffering between different latency data-paths.

\noindent\textbf{Short Streams} For short streams with compile-time known lengths, SMOs pre-reserve sets to avoid allocation and linking overhead. Unknown-length streams use the general storage mechanism.

The {\DLSM} provides local store and load throughput of $>0.9$ tiles per cycle in steady state in all operating modes.





\subsection{Meta Operators}\label{subsec:meta-operators}
Our lightweight meta operators match the STeP operators shown in~\Cref{tab:step-operators} to support the streaming tensor semantics of dynamic workloads and retain logical tile semantics. This includes both control-flow and higher-order operators. Importantly, meta operators are lightweight compared with element-wise or matmul operators, because they operate on the metadata embedded in a tile, amortizing their cost across the tile.

\subsection{Compute}
The fabric's compute resources are divided into tiled matrix-multiply groups (MMGs) and element- or row-wise operator groups (EWGs). Each MMG’s matrix-multiply operator (MMO) performs one fully pipelined transposed matrix multiplication on two input tiles each cycle and has local DLSM memories for its inputs and output to reduce interconnect pressure through hierarchical tiling.

\subsection{Input/Output (IO)}
\name provides a set of IO operators, including linear read/write, dynamic read/write (which fires on a pulse), and random-access read (which accepts a dynamic address of a logical tile). These operators assume that the tensors are stored as physical tiles laid out sequentially in memory. IO operators include programmable address generators that take a base address, output shape, and stride, and produce a sequence of tile addresses.

\subsection{Interconnect}
To support runtime reconfiguration, \name connects {\PSG}s through a packet-switched mesh with deterministic XY routing. During connection setup, the network attempts to allocate a private virtual channel along the stream's route. Each virtual channel has separate buffering within the routers, while sharing physical links. If a private virtual channel is unavailable, the stream uses a shared credit-based fallback channel. In both modes, transfers only occur when the next router has buffer space. A separate control network carries connection setup and credit messages.

\section{Evaluation}\label{sec:eval}

We evaluate \name's performance relative to a static RDA baseline in terms of off-chip traffic, speedup, and area overhead. We also compare the \DLSM against cache baselines. We demonstrate that hardware support for dynamic shapes improves performance, that \name's memory system and scheduler reduce off-chip traffic without programmer effort, and evaluate \name on real-world LLM workloads. Workloads are selected to stress both regular and irregular computation patterns, with varying input sizes to demonstrate performance across diverse scenarios.

\subsection{Experimental Setup}
\begin{table}[t]
\begingroup
  \newcommand{\psgrow}[3]{%
  \multicolumn{2}{>{\raggedright\arraybackslash}%
    p{\dimexpr6.25cm+2\tabcolsep\relax}}{%
    \makebox[1.4cm][l]{#1}\parbox[t]{4.55cm}{#2}\hfill#3}}
  \providecommand{\tblblank}{---}
  \providecommand{\psgline}[2]{\noindent\makebox[\linewidth]{#1\hfill#2}}
  \providecommand{\psgtotal}[1]{\noindent\makebox[\linewidth]{\hfill#1}}
  \caption{\name's evaluated configuration and synthesized area (Synopsys Design Compiler, 16 nm; SMO memories use dual-port SRAMs). Off-chip bandwidth is comparable to two HBM3 stacks~\cite{samsung_hbm3}.}
  \footnotesize
  \centering
  \setlength{\tabcolsep}{1pt}
  \renewcommand{\arraystretch}{1.06}
  \begin{tabular}{@{}>{\raggedright\arraybackslash}p{1cm} >{\raggedright\arraybackslash}p{2.4cm} >{\centering\arraybackslash}p{3.85cm} >{\centering\arraybackslash}p{1cm}@{}}
    \toprule
    \textbf{} & \textbf{Component} & \textbf{Description} & \textbf{Area}\\
    \midrule
    \multirow{2}{*}{\makecell[l]{\textbf{Tile}}}
& Tile size & $16\times16$ BF16 (512\,B/tile) & \tblblank \\      & Metadata & 8\,bit stop \& 1\,bit end token & \tblblank \\
    \midrule
    \multirow{2}{*}{\makecell[l]{\textbf{Peak}}}
      & Off-chip BW & \makecell{1024 B/cycle read\\512 B/cycle write} & \tblblank \\
      & BF16 Peak FLOP & \SI{152}{KFLOPs/cycle} & \tblblank \\
    \midrule
    \multirow{2}{*}{\makecell[l]{\textbf{Mem.}}}
      & SMO cap. & \SI{0.5}{MiB} per SMO (\num{1024} tiles) & \tblblank \\
      & Set size & 8 tiles/set & \tblblank \\
    \toprule
    \multicolumn{4}{@{}l@{}}{%
      \makecell[c]{%
        \footnotesize\itshape
        \textbf{Operators.}
        Area relative to the MO baseline.}}\\[-0.15em]
    \midrule
    \multirow{6}{*}{\makecell[l]{\textbf{{\PSO}s}}}
      & MO & \SI{1}{tile/cycle} & 1.00$\times$ \\
      & IO & \SI{1}{tile/cycle} & 2.95$\times$ \\
      & SMO & \SI{1}{tile/cycle} & 24.07$\times$ \\
      & SMO (local-only) & \SI{1}{tile/cycle} & 20.58$\times$ \\
      & EWO & \SI{256}{FLOPs/cycle} & 10.39$\times$ \\
      & MMO (16$\times$16$\times$16)& \SI{4096}{MACs/cycle} & 39.17$\times$ \\
    \toprule
    \multicolumn{4}{@{}l@{}}{%
      \makecell[c]{%
        \footnotesize\itshape
        \textbf{Fabric}
        Area.}}\\[-0.15em]
    \midrule
    \multirow{5}{*}{\makecell[l]{\textbf{{\PSG}s}}}
      & \psgrow{MMG}{SMO (2) $\to$ MMO $\to$ EWO $\to$ SMO }{\num{16}  } & 48.34\%\\
      & \psgrow{IOG}{IO $\to$ MO (3)}{\num{12}} & 2.29\% \\
      & \psgrow{EWG}{EWO (2) $\to$ MO}{\num{16}  } & 5.36\% \\
      & \psgrow{SMG}{SMO $\to$ MO (3) $\to$ EWO}{\num{36}  } & 28.90\% \\
      & \textbf{Total PSGs} & \makecell{~} & \textbf{84.89\%} \\
    \midrule
    \makecell[l]{\textbf{NoC}} & IC & \makecell{90 transport nodes\\+ neighbor mesh} & 15.11\% \\
    \toprule
    \multicolumn{4}{@{}l@{}}{%
      \makecell[c]{%
        \footnotesize\itshape
        \textbf{Scheduler}
        Area. Area is already included in the PSG area above.}}\\[-0.15em]
    \midrule
    \makecell[l]{\textbf{Sched.}} & Schedulers & \makecell{80 embedded in PSGs} & 1.48\% \\
    \bottomrule
  \end{tabular}
  \label{tab:config-summary}
\endgroup
\end{table}

We implement the \name RDA in Bluespec SystemVerilog~\cite{Nikhil2008bluespec} and simulate with Bluespec's cycle-accurate Bluesim simulator. The evaluated system parameters are listed in~\Cref{tab:config-summary}, and its fabric layout is shown in~\Cref{fig:eval-fabric}. 

For all LLM-layer evaluations, \nameCache uses a 50/50 split between aggregate L1 and L2 capacity: each private L1 holds 256 KiB, and the shared L2 holds 21 MiB, giving a total payload capacity of 42 MiB, matching \name. Both memory systems charge every off-chip read the same fixed latency per tile and draw on the same off-chip bandwidth. Neither system models DRAM access latency which is typically on the order of 100 cycles at 1 GHz. Off-chip traffic therefore costs far less time in our simulations than on hardware. This favors \nameCache, which moves more off-chip data than \name. Because extra traffic is nearly free, the reported speedups over \nameCache are conservative.

We benchmark representative layers with BF16 weights on a single NVIDIA H100 SXM GPU~\cite{nvidia_h100}
with 80 GB HBM3, a 700 W power limit, and an observed SM clock of 1.98 GHz. We use vLLM 0.31.0~\cite{vllm,vllm031_docs} with PyTorch 2.13.0~\cite{pytorch_project} and CUDA 12.9~\cite{cuda129_docs}.
Each layer runs the kernels vLLM dispatches on Hopper: cuBLAS~\cite{cublas129_docs}
for linear projections, FlashAttention-3~\cite{flashattention3_paper,flashattention3_code}
with 16-token KV pages and up to 32 KV splits, Triton~\cite{triton_docs} fused-MoE
kernels, and fused normalization/RoPE kernels adopted from TensorRT-LLM~\cite{tensorrt-dynshape}.
Mamba uses vLLM's causal-convolution and selective-scan kernels for midfill and its
selective-state-update kernel for decode.
Because these are the kernels production deployments run, the baseline reflects the
H100 performance users obtain in practice; we further remove serving overheads by
capturing each layer as a single CUDA graph.

We compute \name and \nameCache roofline attainment as $\max(F/P,\,D_r/B_r,\,D_w/B_w)/t$, using useful tensor FLOPs $F$, useful read bytes $D_r$, useful write bytes $D_w$, and measured latency $t$. \name-Static runs each graph as one spatial phase on a fabric, which is scaled up by increasing the outer array size in~\Cref{fig:eval-fabric} until the program graph fits. Neighbor allocation is disabled and input tensors are padded as is required with a static dataflow accelerator. Since synthesizing each spatial fabric would be very time-intensive, we report its area--delay product (ADP) relative to \name as $(N_{\mathrm{Static}}T_{\mathrm{Static}})/(80\,T_{\name})$, using the PSG count $N$ as an area proxy, and $T$ as the respective layer execution time in cycles.

At 1 GHz, the accelerator admits two 4,096-bit read tiles and one 4,096-bit write tile per cycle, giving independent ceilings of $B_r=1.024$ TB/s and $B_w=0.512$ TB/s. The BF16 peak $P=152$ TFLOP/s is listed in \Cref{tab:config-summary}. For H100, we use $\max(F/P,(D_r+D_w)/B)/t$, with 989.4 TFLOP/s and 3.35 TB/s. All evaluated layers are memory-bound under this accounting.
The time for compiling our examples is dominated by the packing and place \& route algorithms. All of our evaluated examples compile in a few seconds to a few minutes. For evaluations involving the larger {\name}-Static accelerator, we increase the compiler's placement search budget since the designs are otherwise limited by significant interconnect congestion.

\begin{figure}
    \centering
    \input{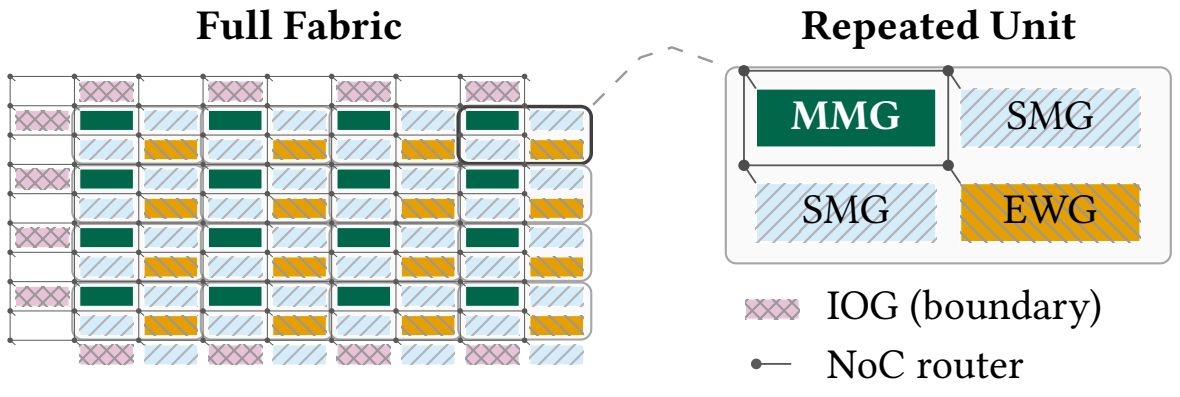}
    \caption{The evaluated fabric comprises a $4\times4$ array of repeated $2\times2$ interior units, with boundary IO. Dots and solid links denote the NoC.}
    \label{fig:eval-fabric}
\end{figure}

\subsection{Shape Dynamism}\label{subsec:eval-dynamic-shapes}

\begin{figure}[t]
\centering
\begingroup
\input{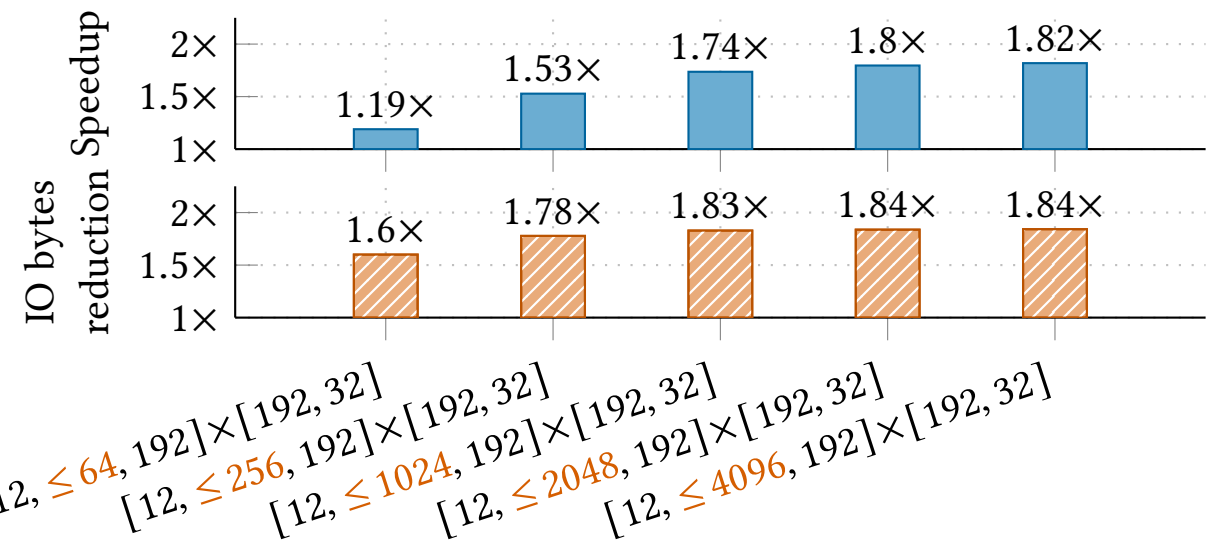}%
\colorlet{raggedcolor}{oiVermillion}

\pgfplotsset{
    sweeplabels/.style={
        xtick={0,1,2,3,4},
        xticklabels={%
            {$[12,{\color{raggedcolor}\leq\!64},192]{\times}[192,32]$},%
            {$[12,{\color{raggedcolor}\leq\!256},192]{\times}[192,32]$},%
            {$[12,{\color{raggedcolor}\leq\!1024},192]{\times}[192,32]$},%
            {$[12,{\color{raggedcolor}\leq\!2048},192]{\times}[192,32]$},%
            {$[12,{\color{raggedcolor}\leq\!4096},192]{\times}[192,32]$}%
        },
        x tick label style={rotate=18, anchor=north east, font=\footnotesize},
    },
}

\begin{tikzpicture}
\footnotesize

\begin{axis}[
    name=spdup,
    eval traffic plot,
    ybar=4pt,
    bar width=12pt,
    width=8.0cm, height=2.45cm,
    xmin=-0.5, xmax=4.5,
    xtick={0,1,2,3,4},
    xticklabels={,,,,},
    ymin=1, ymax=2.25,
    ytick={1,1.5,2},
    yticklabel={\pgfmathprintnumber{\tick}$\times$},
    minor y tick num=0,
    ylabel={Speedup},
    ylabel style={align=center, font=\footnotesize},
    tick label style={font=\footnotesize},
    label style={font=\footnotesize},
    axis lines*=left,
    enlarge x limits=0.08,
    nodes near coords={\pgfmathprintnumber[fixed,precision=2]{\pgfplotspointmeta}$\times$},
    nodes near coords style={font=\footnotesize, anchor=south, inner sep=1pt, yshift=1pt},
]
\addplot[eval bar colors A]
coordinates {(0,1.18957565) (1,1.52808492) (2,1.73636233) (3,1.79600213) (4,1.81862287)};
\end{axis}

\begin{axis}[
    name=io,
    at={(spdup.south west)},
    anchor=north west,
    yshift=-7pt,
    eval traffic plot,
    ybar=4pt,
    bar width=12pt,
    width=8.0cm, height=2.45cm,
    xmin=-0.5, xmax=4.5,
    xtick={0,1,2,3,4},
    sweeplabels,
    ymin=1, ymax=2.25,
    ytick={1,1.5,2},
    yticklabel={\pgfmathprintnumber{\tick}$\times$},
    minor y tick num=0,
    ylabel={\shortstack[c]{IO bytes \\ reduction}},
    ylabel style={align=center, font=\footnotesize},
    tick label style={font=\footnotesize},
    label style={font=\footnotesize},
    axis lines*=left,
    enlarge x limits=0.08,
    nodes near coords={\pgfmathprintnumber[fixed,precision=2]{\pgfplotspointmeta}$\times$},
    nodes near coords style={font=\footnotesize, anchor=south, inner sep=1pt, yshift=1pt},
]
\addplot[eval bar colors B,postaction={pattern=north east lines,pattern color=white}]
    coordinates {(0,1.60000000) (1,1.77777778) (2,1.82857143) (3,1.83732057) (4,1.84172662)};
\end{axis}

\end{tikzpicture}\endgroup
\caption{Ragged linear layer $Y=\mathrm{ReLU}(XW)$ on one GEMM unit, $X\in\mathbb{R}^{12\times M\times192}$ with triangular occupancy along $M$ ({\color[HTML]{B03030}$\leq n$}), and dense weight matrix $W\in\mathbb{R}^{192\times32}$. Speedup (top) and off-chip input and output traffic reduction excluding weights (bottom) over the same STeP program padded to dense. Gains grow with $M$, up to a $1.82\times$ speedup.}
\label{fig:dynamic-shape-small}
\end{figure}

We evaluate \name's support for irregular (ragged) shapes on a linear layer against a padded baseline. Dynamic shape support avoids padding beyond hardware tiles. Speedup grows from $1.19\times$ to $1.82\times$ as $M$ increases (\Cref{fig:dynamic-shape-small}), because hardware tile padding becomes relatively cheaper and the scheduler's fixed overhead is amortized over longer streams. Off-chip traffic falls by up to $1.84\times$, approaching the $2\times$ bound of triangular occupancy as diagonal tile padding becomes negligible.

\subsection{Memory System}

Increasing logical tile size is a standard way to reduce off-chip data movement, and therefore energy consumption, through data reuse. As such, \name's goal is to support larger logical tiles by using on-chip capacity more effectively. We compare \name, \nameLocal, which disables distributed on-chip borrowing, and four cache configurations, which vary the capacity split between L1 and L2, by evaluating batched GEMMs spatially mapped to the fabric in \Cref{fig:eval-fabric}, with on-chip capacity scaled down by $8\times$ from \Cref{tab:config-summary} to keep simulation tractable.

The caches use four-way nonblocking L1s and, where present, a shared L2 organized into four four-way banks. Both levels are write-back caches with eight-tile lines and LRU replacement. Each L1 and L2 bank supports eight outstanding misses with same-line read-miss merging~\cite{800052.801868}. Pipelined L1 accesses and buffered writebacks permit overlap with outstanding transfers. As in \name, accumulation uses the SMO's add-only unit. Both standard and enhanced variants use tile-granular write-validate allocation to avoid fetching old data on L1 write misses. Enhanced variants use deallocation hints naturally embedded in streams to suppress dead-data writebacks.

\begin{figure}[t]
\centering
\begingroup\input{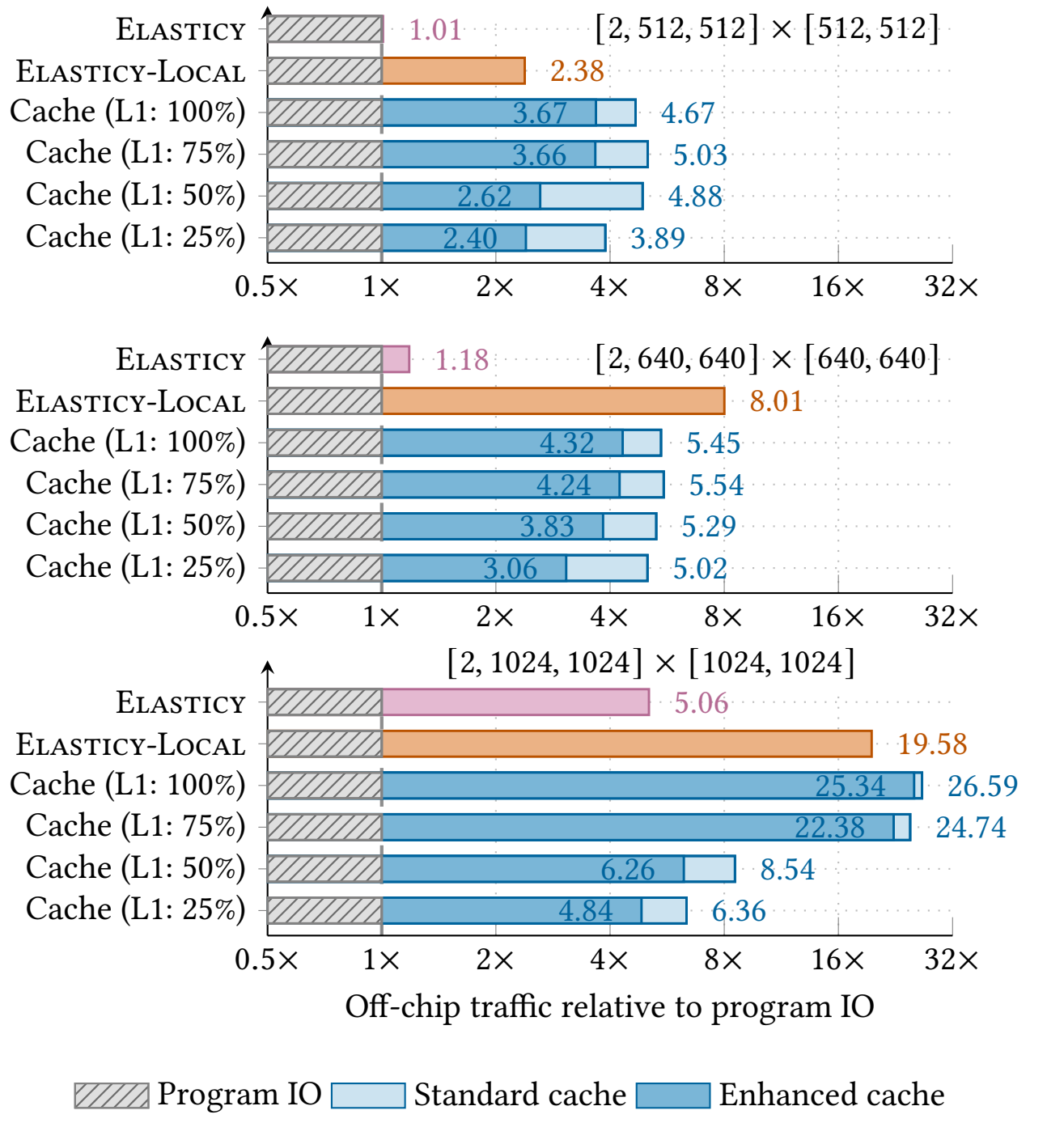}\endgroup
\caption{Off-chip tile transfers for batched GEMMs $[2,N,N]\times[N,N]$ with logical tiling $(N, N/4, N/4)$, normalized to program IO (identical across systems). Enhanced caches add deallocation hints. Through $N=640$, \name stays within $1.2\times$ of program IO, versus $2.4\times$ to $3.1\times$ for the best cache.}
\label{fig:memory-results}
\end{figure}
\begin{figure}[t]
\centering
\begingroup\input{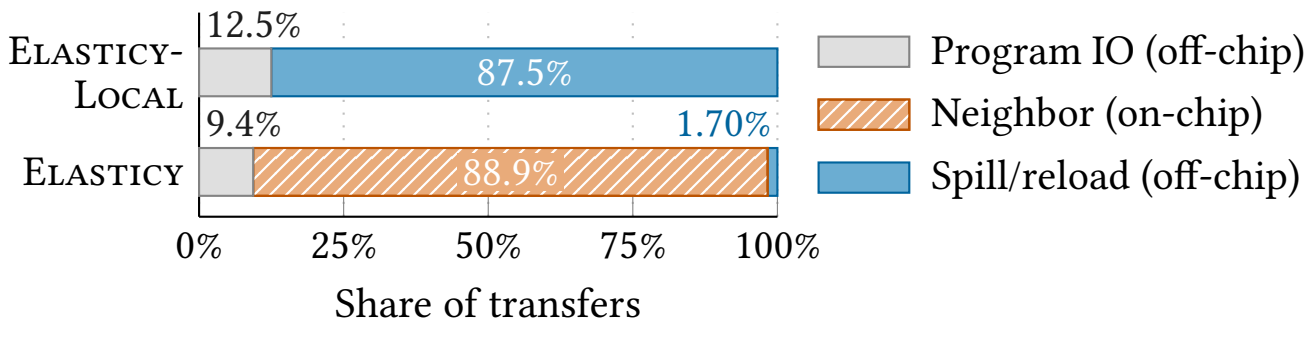}%
\begin{tikzpicture}
\footnotesize
\begin{axis}[
    set layers=standard,
    eval traffic plot,
    xbar stacked,
    bar width=9pt,
    scale only axis,
    width=\dimexpr\columnwidth-4.65cm\relax, height=1.35cm,
    xmin=0, xmax=100,
    xtick={0,25,50,75,100},
    xticklabel={\pgfmathprintnumber{\tick}\%},
    xlabel={Share of transfers},
    ymin=-0.45, ymax=1.6,
    ytick={0,1},
    yticklabels={\name{},{\shortstack[r]{\name{}-\\\textsc{Local}}}},
    tick label style={font=\footnotesize},
    label style={font=\footnotesize},
    axis x line*=bottom,
    axis y line*=left,
    ytick style={draw=none},
    ymajorgrids=false,
    clip=false,
    legend style={
        at={(1.04,0.5)},
        anchor=west,
        font=\footnotesize,
        draw=none,
        fill=none,
        legend columns=1,
        column sep=2pt,
        row sep=1pt,
    },
    legend cell align=left,
]
\addplot[fill=ink!14, draw=ink!55, forget plot] coordinates {(9.3567251462,0) (12.4895920067,1)};
\addplot[eval bar colors B, postaction={pattern=north east lines, pattern color=white}, forget plot] coordinates {(88.9434697856,0) (0,1)};
\addplot[eval bar colors A, forget plot] coordinates {(1.69980506823,0) (87.5104079933,1)};
\addlegendimage{fill=ink!14, draw=ink!55, area legend}
\addlegendentry{Program IO (off-chip)}
\addlegendimage{eval bar colors B, postaction={pattern=north east lines, pattern color=white}, area legend}
\addlegendentry{Neighbor (on-chip)}
\addlegendimage{eval bar colors A, area legend}
\addlegendentry{Spill/reload (off-chip)}
\pgfplotsextra{\begin{pgfonlayer}{axis foreground}
\node[font=\footnotesize,anchor=south west,inner sep=1pt,text=ink] at (axis cs:0,1.29) {12.5\%};
\node[font=\footnotesize,text=white] at (axis cs:56.2447960034,1) {87.5\%};
\node[font=\footnotesize,anchor=south west,inner sep=1pt,text=ink] at (axis cs:0,0.29) {9.4\%};
\node[font=\footnotesize,text=white,fill=oiVermillion!52,inner sep=1pt] at (axis cs:53.828460039,0) {88.9\%};
\node[font=\footnotesize,anchor=south east,inner sep=1pt,text=oiBlue!88!black] at (axis cs:100,0.29) {1.70\%};
\end{pgfonlayer}}
\end{axis}
\end{tikzpicture}\endgroup
\caption{Data movement when enabling \name's distributed neighbor sharing for $[2,640,640]\times[640,640]$.}
\label{fig:traffic-breakdown}
\end{figure}

At $N=512$ and $N=640$, \name needs little traffic beyond program IO, $58.0\%$ and $61.4\%$ less than the best cache (\Cref{fig:memory-results}). Deallocation hints reduce traffic for every cache configuration, yet \name still outperforms the best enhanced cache, indicating \name's benefits extend beyond suppressing dead-data writebacks.  \nameLocal, which cannot borrow neighbor capacity, transfers up to $8\times$ program IO, further indicating \name's savings come from capacity sharing.

At $N=640$, \name replaces \nameLocal's spills with neighbor transfers that stay on-chip, saving off-chip bandwidth and energy (\Cref{fig:traffic-breakdown}). Neighbor transfers account for $88.9\%$ of \name's data movement, while off-chip spills and reloads account for $87.5\%$ in \nameLocal. Overall, \name reduces total off-chip traffic by 85\% relative to \nameLocal.

At size 1024, \name's total off-chip traffic rises to $5.06\times$ program IO, which is within 5\% of the best cache configuration at $4.84\times$ program IO. The logical tiles are too large for either memory system, and such a program should be recompiled with smaller tiling. However, we note that the current \name implementation leaves spilled data off-chip after a read, so repeated reads incur repeated off-chip transfers. Restoring such data to on-chip storage when space becomes available is an opportunity for further improvement. \name uses virtual pointers and supports data eviction to enable this type of future optimization. 

Cache behavior strongly depends on the capacity split. Configurations retaining more shared L2 capacity reduce off-chip traffic relative to \nameLocal, as expected given the larger shared capacity. However, the shared pool of memory comes at a cost: at $N=512$, the 25\% L1 is 66.8\% slower than the 50\% L1 despite less off-chip traffic. Although this greater pool of shared memory enables additional tile reuse for large GEMMs, the inverse is also true. For small GEMMs, a larger L1 would enable all reused data to be retained locally. This highlights a fundamental problem; the tiered capacity split is fixed at design time. In contrast, \name explicitly stores data based on known lifetimes and coordinates use of distributed on-chip capacity in all cases.

The off-chip traffic savings translate into $2.40\times$, $2.98\times$, and $1.96\times$ end-to-end speedups over the fastest evaluated cache configuration at N=512, 640, and 1024, respectively. Speedup over the cache at $N=1024$, despite less off-chip traffic, comes from L2 access times. 

With statically partitioned scratchpads and no automatic spilling, buffers exceeding their assigned capacity would require compiler-managed transfers or changes to tiling and buffering. \name handles this overflow through neighboring storage and off-chip spilling while preserving the chosen mapping.

\subsection{Dynamic Spilling Between Kernel Runs}\label{subsec:kernel-on-chip}

We next evaluate how \name's fine-grained scheduler and dynamic memory system reduce off-chip memory traffic between consecutive, dependent kernel executions. Here, a kernel is defined as a program that executes to completion as one bulk-synchronous invocation. In many spatial accelerators, intermediate results must be written to off-chip memory when time-multiplexing the fabric between kernels, incurring bandwidth overhead. If the intermediate results are shared between kernels, this results in unnecessary memory traffic. \name's \DLSM retains intermediate results on-chip between kernels when capacity permits.

Retaining intermediates in the \DLSM reduces off-chip writes by up to $2\times$ and reads by up to $1.33\times$ on linear layers at batch 32 (\Cref{fig:kernel-over-kernel-high-batch}). Larger batches produce more intermediates, but \name still reduces traffic by deciding which tiles to keep on chip and which to spill based on available capacity.

\begin{figure}[t]
\centering
\begingroup
\input{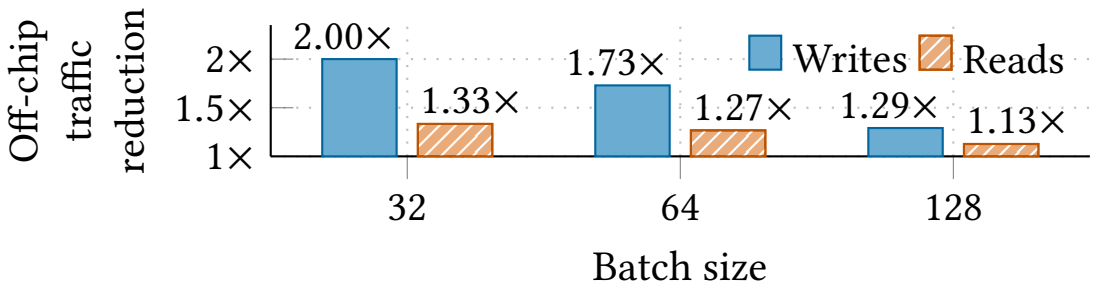}%

\begin{tikzpicture}
\footnotesize
\begin{axis}[
    name=ax1,
    eval traffic plot,
        tick label style={font=\footnotesize},
        label style={font=\footnotesize},
    width=7cm, height=2.45cm,
    xlabel={Batch size},
    ylabel={Off-chip\\traffic\\reduction},
    ylabel style={align=center},
    ybar=4pt,
    single ybar legend,
    bar width=14pt,
    symbolic x coords={32,64,128},
    xtick=data,
    axis lines*=left,
    enlarge x limits=0.25,
    ymin=1, ymax=2.35,
    ytick={1,1.5,2},
    yticklabel={\pgfmathprintnumber{\tick}$\times$},
    legend style={at={(0.99,0.99)}, anchor=north east, inner sep=1pt, font=\footnotesize, legend columns=2},
    legend cell align=left,
    legend style={draw=none,fill=none},
    nodes near coords={\pgfmathprintnumber[fixed,fixed zerofill,precision=2]{\pgfplotspointmeta}$\times$},
    nodes near coords style={font=\footnotesize},
    every node near coord/.append style={yshift=-1pt},
]

\addplot[eval bar colors A, every node near coord/.append style={xshift=-3pt}]
    coordinates {(32,2) (64,1.72972972972973) (128,1.29129886506936)};
\addlegendentry{Writes}

\addplot[eval bar colors B, postaction={pattern=north east lines, pattern color=white}, every node near coord/.append style={xshift=3pt}]
    coordinates {(32,1.33333333333333) (64,1.26732673267327) (128,1.12713263621354)};
\addlegendentry{Reads}

\end{axis}

\end{tikzpicture}%
\endgroup
\caption{Off-chip traffic reduction while serially executing two linear layers of shape $[b, 128, 128] \times [b, 128, 128]$, by retaining intermediates on-chip during reconfiguration. Writes fall by up to $2\times$ and reads by up to $1.33\times$ at $b=32$.}
\label{fig:kernel-over-kernel-high-batch}
\end{figure}

\subsection{Overlapping Kernel Execution}\label{subsec:eval-overlap}

\name's fine-grained scheduler enables partial overlap of consecutive kernels.
We compile pairs of kernels in two configurations: (1)~separate bitstreams, where each kernel runs independently and the total runtime is the sum of their runtimes, and (2)~a combined bitstream, where both kernels are co-compiled so the scheduler can reassign resources freed by kernel~A to kernel~B before A~completes. This experiment isolates the scheduler’s contribution from dynamic-shape and memory-capacity effects: both configurations transfer identical amounts of data for each kernel pair, with no intermediate spilling. We evaluate two kernel pairs with batch size~8 and 128$\times$128 inputs and weights. In the notation $K_1 \to K_2$, each side denotes a kernel mapped to the fabric in sequence. We generously assume that the reference architecture has no kernel launch overhead beyond the bulk-synchronous barrier.

\begin{figure}[t]
\centering
\begingroup\begin{tikzpicture}[scale=2.4, yscale=.5, font=\footnotesize]
\fill[gray!8] (0.550,0.0000) rectangle (2.750,-0.6000);
\fill[color=oiBlue,opacity=0.85] (0.550474,-0.0100) rectangle (1.277117,-0.0900);
\fill[color=oiBlue,opacity=0.85] (0.595837,-0.5100) rectangle (1.286517,-0.5900);
\fill[color=oiBlue,opacity=0.85] (1.286862,-0.1100) rectangle (1.531678,-0.1900);
\fill[color=oiBlue,opacity=0.85] (1.298267,-0.2100) rectangle (2.005549,-0.2900);
\fill[color=oiBlue,opacity=0.85] (1.324290,-0.4100) rectangle (2.014130,-0.4900);
\draw[gray!60,dotted,line width=0.6pt] (2.014194,0.0000) -- (2.014194,-0.6000);
\fill[color=oiGreen,opacity=0.85,postaction={pattern=north east lines,pattern color=white}] (2.014669,-0.0100) rectangle (2.741311,-0.0900);
\fill[color=oiGreen,opacity=0.85,postaction={pattern=north east lines,pattern color=white}] (2.060096,-0.3100) rectangle (2.749935,-0.3900);
\draw[color=oiVermillion,dashed,line width=0.5pt] (2.750000,0.0000) -- (2.750000,-0.6000);
\draw[gray!40,line width=0.2pt] (0.550,-0.1000) -- (2.750,-0.1000);
\draw[gray!40,line width=0.2pt] (0.550,-0.2000) -- (2.750,-0.2000);
\draw[gray!40,line width=0.2pt] (0.550,-0.3000) -- (2.750,-0.3000);
\draw[gray!40,line width=0.2pt] (0.550,-0.4000) -- (2.750,-0.4000);
\draw[gray!40,line width=0.2pt] (0.550,-0.5000) -- (2.750,-0.5000);
\draw[color=oiGreen,line width=0.5pt] (0.490,0.0000) -- (0.490,-0.3000);
\node[anchor=east,color=oiGreen] at (0.455,-0.1500) {MMO};
\draw[color=oiVermillion,line width=0.5pt] (0.490,-0.3000) -- (0.490,-0.6000);
\node[anchor=east,color=oiVermillion] at (0.455,-0.4500) {SMO/IO};
\node[anchor=west] at (2.780,-0.3000) {Unfused};
\draw[gray!60,line width=0.4pt] (0.550,0.0000) rectangle (2.750,-0.6000);
\fill[gray!8] (0.550,-0.7000) rectangle (2.750,-1.3000);
\fill[color=oiBlue,opacity=0.85] (0.550474,-0.7100) rectangle (1.277117,-0.7900);
\fill[color=oiGreen,opacity=0.85,postaction={pattern=north east lines,pattern color=white}] (1.287293,-0.7100) rectangle (1.993367,-0.7900);
\fill[color=oiBlue,opacity=0.85] (0.595837,-1.2100) rectangle (1.286517,-1.2900);
\fill[color=oiGreen,opacity=0.85,postaction={pattern=north east lines,pattern color=white}] (1.311290,-1.0100) rectangle (2.001991,-1.0900);
\fill[color=oiBlue,opacity=0.85] (1.286862,-0.8100) rectangle (2.002768,-0.8900);
\fill[color=oiBlue,opacity=0.85] (1.298267,-0.9100) rectangle (2.088297,-0.9900);
\fill[color=oiBlue,opacity=0.85] (1.324290,-1.1100) rectangle (2.096878,-1.1900);
\draw[color=oiVermillion,dashed,line width=0.5pt] (2.096942,-0.7000) -- (2.096942,-1.3000);
\draw[gray!40,line width=0.2pt] (0.550,-0.8000) -- (2.750,-0.8000);
\draw[gray!40,line width=0.2pt] (0.550,-0.9000) -- (2.750,-0.9000);
\draw[gray!40,line width=0.2pt] (0.550,-1.0000) -- (2.750,-1.0000);
\draw[gray!40,line width=0.2pt] (0.550,-1.1000) -- (2.750,-1.1000);
\draw[gray!40,line width=0.2pt] (0.550,-1.2000) -- (2.750,-1.2000);
\draw[color=oiGreen,line width=0.5pt] (0.490,-0.7000) -- (0.490,-1.0000);
\node[anchor=east,color=oiGreen] at (0.455,-0.8500) {MMO};
\draw[color=oiVermillion,line width=0.5pt] (0.490,-1.0000) -- (0.490,-1.3000);
\node[anchor=east,color=oiVermillion] at (0.455,-1.1500) {SMO/IO};
\node[anchor=west] at (2.780,-1.0000) {Fused};
\draw[gray!60,line width=0.4pt] (0.550,-0.7000) rectangle (2.750,-1.3000);
\node[anchor=west,color=oiVermillion] at (2.725,-1.2250) {1.42$\times$};
\end{tikzpicture}\endgroup
\caption{Timelines of \texttt{square(lin(x))},$\to$,\texttt{lin(x)} with separate bitstreams (\textbf{unfused}, a bulk-synchronous kernel boundary) and one combined bitstream (\textbf{fused}). Solid blue and hatched green segments are groups of the first and second kernel. Fusing gives a $1.42\times$ speedup.}
\label{fig:overlap-kernel-swimlane}
\end{figure}
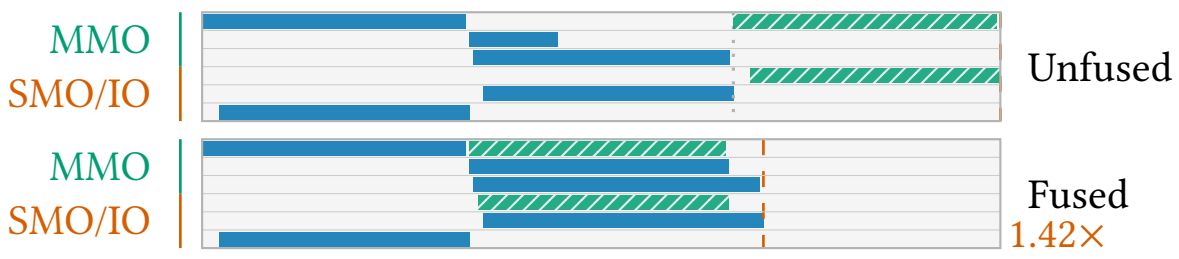

Fusion speeds up \texttt{square(lin(x))}\,$\to$\,\texttt{lin(x)} by $1.42\times$ (\Cref{fig:overlap-kernel-swimlane}) and two sequential \texttt{lin(lin(x))}\,$\to$\,\texttt{lin(lin(x))} layers, a common pattern, by $1.35\times$. The degree of overlap depends on the temporal spread among {\VSG}s within each kernel: composed multi-stage kernels contain groups with staggered lifetimes that free resources progressively, creating windows for the next kernel's groups to begin. For graphs with deep pipelines, kernel completion times can vary widely, making bulk-synchronous execution on static RDAs particularly costly; \name reduces this overhead by allowing subsequent kernels to begin as resources become available. Unlike prior RDA work on kernel looping~\cite{koeplinger2024kernelloopingeliminatingsynchronization}, \name can also overlap different kernels.

\subsection{Performance during LLM Decode}

We evaluate \name on selected LLM layers described in \Cref{tab:llm-architectures}. \name{} implements its attention kernels using flash decode~\cite{dao2022flashattention}. \name achieves favorable memory-roofline attainment across all four examples (\Cref{tab:llm-eval}).
\begin{table}[t]
\centering
\caption{Architecture of the selected LLM layers.}
\label{tab:llm-architectures}
\footnotesize
\setlength{\tabcolsep}{4pt}
\begin{tabular}{@{}p{.22\linewidth}p{\dimexpr.78\linewidth-8pt\relax}@{}}
\toprule
\textbf{Model} & \textbf{Layer architecture} \\
\midrule
\raggedright\textbf{Llama 3--8B}~\cite{hf_llama3_8b} &
\raggedright Dense Transformer; grouped-query attention (GQA) with 32 query and 8 KV heads; dense feed-forward network. \tabularnewline
\raggedright\textbf{Qwen1.5-MoE-A2.7B}~\cite{hf_qwen15_moe} &
\raggedright Transformer; multi-head attention with 16 query and 16 KV heads; mixture-of-experts feed-forward network with data-dependent expert dispatch and recombination. \tabularnewline
\raggedright\textbf{DFlash (Qwen3-4B)}~\cite{hf_dflash_qwen3,hf_qwen3_4b} &
\raggedright Speculative decoding draft model; GQA with 32 query and 8 KV heads. \tabularnewline
\raggedright\textbf{Mamba-1.4B}~\cite{hf_mamba14b} &
\raggedright State space model with convolution and recurrent state updates. \tabularnewline
\bottomrule
\end{tabular}
\end{table}

\begin{table}[t]
\centering
\caption{Roofline attainment, speedup, and Area-Delay-Product (ADP) of single LLM layers during decode. Attainment is normalized to each platform's memory ceiling. Llama and Qwen decode one position for 16 requests with ragged context lengths, which \name-Static pads to the longest. Mamba decodes one position with carried convolution and recurrent state.}
\label{tab:llm-eval}
\footnotesize
\setlength{\tabcolsep}{1.5pt}
\begin{threeparttable}
\begin{tabular*}{\columnwidth}{@{\extracolsep{\fill}}lccccccc@{}}
\toprule
\textbf{Model} & \textbf{Batch} & \textbf{ISL} & \textbf{OSL} & \multicolumn{2}{c}{\textbf{Roofline (\%)}} & \multicolumn{2}{c}{\textbf{vs. \name-Static}} \\
\cmidrule(lr){5-6}\cmidrule(l){7-8}
 & & & & \textbf{\name} & \textbf{H100} & \textbf{Speedup} & \textbf{ADP gain} \\
\midrule
\textbf{Llama 3--8B} & 16 & 128--1920\tnote{\dag} & \phantom{0}1 & \textbf{92.59} & 69.67 & 1.15$\times$ & ~6.34$\times$ \\
\textbf{DFlash} & \phantom{0}1 & \phantom{0}256 & 16 & \textbf{96.50} & 51.54 & 1.00$\times$ & 15.23$\times$ \\
\textbf{Mamba-1.4B} & \phantom{0}1 & \phantom{0}\textit{N/A} & \phantom{0}1 & \textbf{82.32} & 36.14 & 0.99$\times$ & ~9.03$\times$ \\
\textbf{Qwen MoE} & 16 & 128--1920\tnote{\dag} & \phantom{0}1 & \textbf{75.23} & 73.06 & 1.14$\times$ & 12.00$\times$ \\
\bottomrule
\end{tabular*}
\begin{tablenotes}[flushleft]
\footnotesize
\item[$\dag$] Ragged context lengths; mean 1024.
\end{tablenotes}
\end{threeparttable}
\end{table}


\begin{figure}[t]
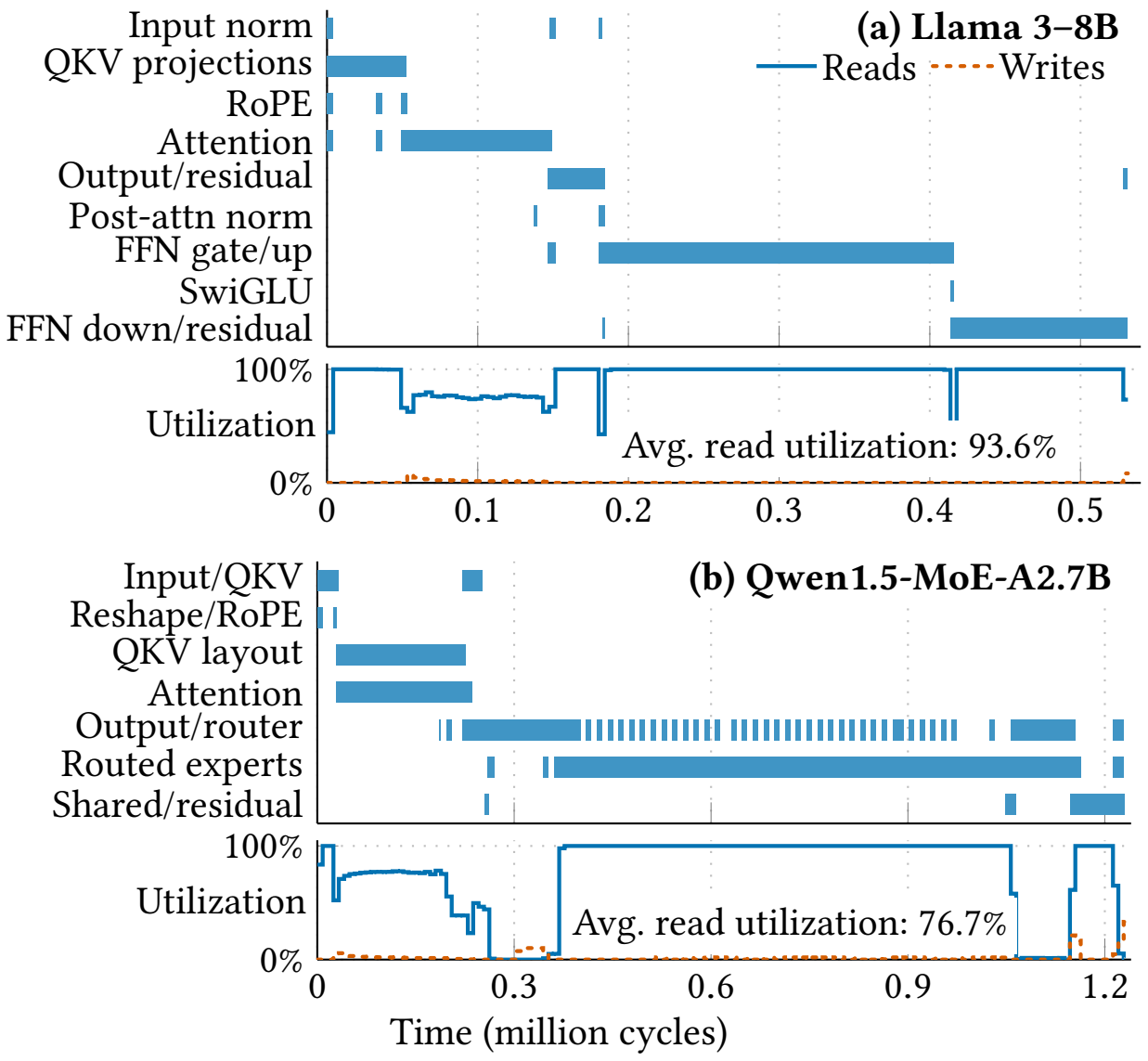

\centering
\captionsetup[subfigure]{hypcap=true}
\begin{subfigure}[t]{\columnwidth}
  \phantomsubcaption\label{fig:llama-bandwidth}
  \makebox[\linewidth][r]{\input{replacement-figures/tikz/models/llama_bandwidth.tex}}
\end{subfigure}
\par\smallskip
\begin{subfigure}[t]{\columnwidth}
  \phantomsubcaption\label{fig:qwen-bandwidth}
  \makebox[\linewidth][r]{\input{replacement-figures/tikz/models/qwen_bandwidth.tex}}
\end{subfigure}
\par\centering{\small Time (million cycles)}
\caption{Memory-bandwidth utilization on (a) Llama and (b) Qwen decode. Utilization includes all transfers (tile padding, control traffic, and final drain), not just useful bytes used for roofline attainment. The upper rows show when each phase is active. Llama's reads are near saturation outside of attention; Qwen's reads dip during expert dispatch and recombination.}
\label{fig:model-bandwidth}
\end{figure}

\textbf{Llama.} \name reaches 92.6\% of its memory roofline, against 69.7\% on H100. The dip during attention in \Cref{fig:llama-bandwidth} comes from interconnect congestion. \name's dynamic execution model supports ragged tensors including KV caches, which avoids padded reads and computation in attention. To quantify this benefit, we ran the same Llama model with padded shapes, holding placement and encoded instructions fixed. This change affects only attention, which runs $1.85\times$ faster with ragged execution. Across the whole layer, ragged-shape support results in a $1.15\times$ speedup and reduces off-chip read traffic by 11.6\%.

\textbf{DFlash.} \name reaches 96.5\% of its memory roofline, against 51.5\% on H100. On H100, PyTorch Profiler shows the matching cached-decode layer runs as 15 separate kernels; attention, normalization, rotary embedding, cache writes, and activation take about 24\% of the profiled layer span, and another 5\% is spent between kernels. These costs are poorly amortized over the 16-position draft block. \name instead fuses the projections and auxiliary operations through a streaming dataflow graph, amortizing scheduling overhead at stream boundaries.

\textbf{Mamba.} \name reaches 82.3\% of its memory roofline, against 36.1\% on H100. \name combines normalization and input projection in one spatial dataflow phase and output projection and residual addition in another, reducing the execution boundaries between dependent operations. On H100 these run as separate kernels: normalization, convolution, state-update, and residual kernels take about 18\% of layer latency, and another 9\% is spent between kernels despite CUDA-graph execution. Batch one also leaves H100's projections little parallelism: the evaluated cuBLAS mapping launches only 12 thread blocks for one projection on 132 SMs.

\textbf{Qwen.} \name reaches 75.2\% of its memory roofline, against 73.1\% on H100. Intermediate buffers start to spill, but spill writes and reload reads account for only 0.80\% of total off-chip traffic; the performance degradation comes from the utilization drops in \Cref{fig:qwen-bandwidth}. Most of the Qwen decode runtime (61\%) happens in the \texttt{fused\_moe\_kernel}, which runs in just two launches and reaches about 90\% of HBM peak, so per-kernel fixed costs are amortized over the larger transfers.


\Cref{fig:qwen-bandwidth} shows additional utilization drops in \name when transforming 2D tiles into vectors for expert dispatch and converting the results back. The current memory organization addresses whole tiles, requiring these transformations to process one row or column at a time. This serialization adds latency and leaves the memory bandwidth underutilized. Parallelizing these transformations to achieve a throughput of one full tile per cycle would require costly hardware support: either independently addressed vector-level memory banks that allow concurrent access to different rows across multiple tiles, or dedicated transpose buffers. We believe this is a fundamental cost of tile-based hardware architectures.

\begin{table*}[t]
\centering
\caption{Roofline attainment, speedup, and Area-Delay-Product (ADP) of single LLM layers during midfill. Slash-separated lengths describe separate requests within the batch. \textsc{Cache} and \textsc{Static} abbreviate \nameCache and \name-Static. }
\label{tab:midfill-eval}
\footnotesize
\setlength{\tabcolsep}{3.5pt}
\begin{threeparttable}
\begin{tabular}{lcccccccccc}
\toprule
\textbf{Model} & \textbf{Batch} & \textbf{KV\$ ISL} & \textbf{Midfill input} & \multicolumn{4}{c}{\textbf{Roofline attainment (\%)}} & \multicolumn{2}{c}{\textbf{Speedup over}} & \textbf{ADP improvement} \\
\cmidrule(lr){5-8}\cmidrule(lr){9-10}
 & & \textbf{(tokens)} & \textbf{(tokens)} & \textbf{\name} & \textbf{\textsc{Cache}} & \textbf{\textsc{Static}} & \textbf{H100} & \textbf{\textsc{Cache}} & \textbf{\textsc{Static}} & \textbf{over \textsc{Static}} \\
\midrule
\textbf{Llama 3-8B} & 2 & 1024 / 2048 & 32 / 16 & 80.8 & 61.7 & \textbf{82.7} & 65.0 & 1.31$\times$ & 0.98$\times$ & 22.58$\times$ \\
\textbf{DFlash} & 2 & 1024 / 1024 & 32 / 32 & 76.0 & 48.3 & \textbf{77.7} & 50.6 & 1.58$\times$ & 0.98$\times$ & 20.54$\times$ \\
\textbf{Mamba-1.4B} & 1 & \textit{N/A} & 16 & \textbf{76.8} & 74.9 & 73.8 & 18.3 & 1.03$\times$ & 1.04$\times$ & 21.85$\times$ \\
\textbf{Qwen MoE} & 2 & 1024 / 2048 & 32 / 32 & 87.5 & 69.6 & \textbf{89.8} & 71.4 & 1.26$\times$ & 0.97$\times$ & 31.67$\times$ \\
\bottomrule
\end{tabular}
\end{threeparttable}
\end{table*}

\textbf{Folding.} To evaluate the benefit of \name's phase-switching, we compare the folded layers against fully spatially mapped and padded versions of the same computation graphs, preserving global memory bandwidth. Folding the graph in time costs \name at most a 1\% performance hit and drastically reduces required fabric size, with 5.5$\times$ to 15.3$\times$ fewer PSGs. On Qwen and Llama, \name is faster than \name-Static because \name-Static pads their ragged batches to the longest sequence. The extra FLOPS provided by the additional PSGs do not improve performance because the layers remain memory bound. \name therefore improves the ADP by 6.3$\times$ on Llama to 15.2$\times$ on DFlash, whose graph requires the largest static fabric. Supporting spatial execution and temporal folding gives \name a broader mapping space than spatial-only RDAs. When graph dependencies limit effective pipeline parallelism, folding allows fabric reuse, improving utilization and area efficiency. This further allows for operations in the active phase to use proportionately more of the total PSGs, reducing pressure on the global interconnect.

\subsection{Performance during Midfill}\label{subsec:eval-midfill}
Midfill processes a block of new tokens against cached context. This is often seen in agentic workloads or turn-based chatbots. Midfill increases weight reuse and enlarges intermediate buffers. Therefore, \name's performance depends on how efficiently the memory system can handle dynamically sized intermediates. For each workload, we report the mapping with the highest \name roofline attainment (\Cref{tab:midfill-eval}). All workloads remain memory-bound.

\textbf{Llama.} \name reaches 80.8\% of its memory roofline, compared with 65.0\% on H100, and is $1.31\times$ faster than \nameCache despite negligible additional DRAM traffic in the cache baseline. However, the 48-position, 4,096-channel hidden activation occupies 384\,KiB: larger than a 256\,KiB cache L1 but within a 512\,KiB \DLSM local store. Therefore \name's speedups over \nameCache are due to L2 access times, demonstrating that equivalent aggregate SRAM capacity does not result in equivalent performance. 

\textbf{DFlash.} Midfill increases the intermediate activations over the L1 capacity. \name achieves a $1.58\times$ speedup over \nameCache, with nearly identical off-chip traffic. Like Llama, this result demonstrates that avoiding DRAM spills is insufficient to sustain performance due to L2 access times.

\textbf{Mamba.} A 16-position continuation with carried convolution and recurrent state reaches 76.8\% of the memory roofline on \name and 74.9\% on \nameCache, giving a $1.03\times$ speedup. Mamba’s carried recurrent state has a fixed size independent of prefix length. This prevents intermediate storage from growing with prefix length, allowing both memory systems to achieve similar performance. However, longer midfill blocks will still increase intermediate activation storage, creating additional memory pressure. 

\textbf{Qwen.} Unlike Llama and DFlash, Qwen exposes substantial differences in off-chip traffic. Relative to \nameCache, \name reduces intermediate off-chip traffic by 93.6\% and total off-chip traffic by 10.8\%, achieving a $1.26\times$ speedup. This extends the GEMM experiments’ finding to a full MoE layer: retaining intermediates on-chip reduces backing-memory traffic and improves latency. 

\textbf{H100.} Across the evaluated midfill workloads, the H100's roofline attainment is limited by kernels that do not reach peak bandwidth and latencies associated with kernel-by-kernel execution. Llama and Qwen are limited by their attention and routing and state update phases. For Qwen, expert GEMMs run in a fused kernel that reaches high memory bandwidth utilization, but kernel dispatch delays execution. Mamba performs especially poorly on the H100. The 16-token input does not fill the scan kernel's chunk size, and Mamba spends 35\% of its execution time on scan. 

\textbf{Static RDA.} As in decode, \name achieves latency comparable to that of a static RDA, despite folding to a substantially reduced fabric size. ADP improves by more than $20\times$ across all evaluated workloads, mainly driven by resource requirements of IO {\PSG}s. Just like in decode, additional compute units do not help increase performance because the workload is memory bound.

\subsection{Area Overhead}

Supporting dynamism costs little area. We synthesize \name using the Synopsys Design Compiler with a 16 nm technology library; each SMO memory is implemented using dual-port SRAMs. Results are presented in~\Cref{tab:config-summary}. The three components that enable dynamism are all small. The 80 schedulers occupy 1.48\% of the total area. Meta operators operate only on the stop tokens embedded in tiles, so each is $39.17\times$ smaller than MMOs and $10.39\times$ smaller than EWOs. In total, MOs occupy only 2.94\% of the fabric area. Synthesizing \nameLocal isolates the \DLSM: distributed communication adds 16.94\% to SMO area and 7.70\% to total fabric area. Together, these dynamic features account for under 12\% of the fabric. The interconnect's 15.11\% is common for RDAs, which often devote substantial area to communication~\cite{prabhakar2018plasticine}.

\subsection{Power}
\begin{table}[!htbp]
\begingroup
\caption{\name's synthesized power at 1 GHz using a 16\,nm technology library.}
\centering
\footnotesize
\setlength{\tabcolsep}{6pt}
\begin{tabular}{lrrr}
\toprule
\textbf{Block} & \textbf{Count} & \textbf{Power (W)} & \textbf{Share} \\
\midrule
\textbf{MMG} & 16 & 100.61 & 41.49\% \\
\textbf{SMG} & 36 & 70.39 & 29.03\% \\
\textbf{EWG} & 16 & 15.77 & 6.50\% \\
\textbf{IOG} & 12 & 5.95 & 2.46\% \\
\textbf{PSG total} & 80 & 192.72 & 79.47\% \\
\textbf{Interconnect} & -- & 49.77 & 20.53\% \\
\textbf{Total} & -- & 242.49 & 100.00\% \\
\toprule
\multicolumn{4}{@{}l@{}}{%
  \makecell[c]{%
    \footnotesize\itshape
    \textbf{Scheduler}
    Power. Power is already included in the PSG power above.}}\\[-0.15em]
\midrule
\textbf{Schedulers} & 80 & 1.12 & 0.46\% \\
\bottomrule
\end{tabular}
\label{tab:power-breakdown}
\endgroup
\end{table}

\name delivers 152 TFLOPS of BF16 compute at 242.49 W in 16 nm at 1 GHz, or 0.63 TFLOPS/W. The H100-SXM delivers 989.4 TFLOPS of dense BF16 at 590 W for the die alone (excluding about 110 W of HBM power) in 4 nm at 1.98 GHz~\cite{nvidia_h100}, or 1.68 TFLOPS/W. As a first-order estimate at the same node, scaling \name's power linearly from 16 nm to 4 nm gives 60.62 W and 2.51 TFLOPS/W, $1.50\times$ the efficiency of the H100-SXM, despite providing $2.36\times$ as much SRAM per peak BF16 FLOP/s.

\section{Conclusion}\label{sec:conclusion}
This paper presents \name, a runtime-reconfigurable dataflow accelerator for dense workloads with coarse-grained dynamism. By combining stream-level runtime scheduling, lightweight rank-aware control, and a distributed linked streaming memory, \name shows that dynamic capabilities can be added with minimal overhead while delivering substantial performance and memory benefits on modern AI workloads.

\section{Acknowledgments}\label{sec:acknowledgments}
Generative AI has been used to aid in implementation and testing of \name. All thoughts and ideas remain ours.

\clearpage
\bibliographystyle{ACM-Reference-Format}
\bibliography{refs}

\clearpage
\appendix

\clearpage

\end{document}